\documentclass[10pt]{article} 
\usepackage[utf8]{inputenc} 

\usepackage{geometry} 
\usepackage{graphicx} 
\usepackage[parfill]{parskip} 
\usepackage{color}
\usepackage{booktabs} 
\usepackage{array} 
\usepackage{paralist} 
\usepackage{verbatim} 
\usepackage{subfig} 
\usepackage{authblk}
\usepackage{bbding}
\usepackage{amssymb,amsmath}

\usepackage{fancyhdr} 
\usepackage{sectsty}
\allsectionsfont{\sffamily\mdseries\upshape} 

\usepackage[nottoc,notlof,notlot]{tocbibind} 
\usepackage[titles,subfigure]{tocloft} 

\newcommand{\bi}{\begin{itemize}}
\newcommand{\ei}{\end{itemize}}
\newcommand{\ben}{\begin{enumerate}}
\newcommand{\een}{\end{enumerate}}
\newcommand{\be}{\begin{equation}}
\newcommand{\ee}{\end{equation}}
\newcommand{\bea}{\begin{eqnarray}}
\newcommand{\eea}{\end{eqnarray}}
\newcommand{\nn}{\nonumber}

\def\bbaa{b\bar b \gamma\gamma}
\def\bb{b\bar b}
\def\tt{t\bar t}
\def\tautau{\tau^+\tau^-}
\def\bbll{b\bar b \tau^+\tau^-}
\def\cba{\cos(\beta-\alpha)}
\def\cbashort{c_{\beta-\alpha}}
\def\sba{\sin(\beta-\alpha)}
\def\tb{\tan\beta}

\def\lhhh{\lambda^{hhh}}
\def\lhhH{\lambda^{hhH}}
\def\lhHH{\lambda^{hHH}}

\def\iab{{\rm ab}^{-1}}

\title{Measuring the 2HDM Scalar Potential at LHC14}
\author[1]{Vernon Barger \thanks{barger@physics.wisc.edu}}
\author[1]{Lisa L.~Everett \thanks{leverett@wisc.edu}}
\author[2]{Chris~B.~Jackson \thanks{chris@uta.edu}}
\author[1]{Andrea D.~Peterson \thanks{adpeterson2@wisc.edu}}
\author[1]{Gabe Shaughnessy \thanks{gshau@hep.wisc.edu}}
\affil[1]{Department of Physics, University of Wisconsin, Madison, WI 53706, USA}
\affil[2]{Department of Physics, University of Texas at Arlington, Arlington, TX 76019, USA}

\date{} 

\begin{document}

\maketitle

\begin{abstract}
After the extraordinary discovery of the Higgs boson at the LHC, the next goal is to pin down its underlying dynamics by 
measuring the Higgs self-couplings, along with its couplings to gauge and matter particles.  As a prototype model of new physics in the scalar sector, we consider the Two Higgs Doublet Model (2HDM) with CP-conservation, 
and evaluate the prospects for measuring the trilinear scalar couplings among the CP-even Higgs bosons $h$ and $H$ ($\lhhh$, $\lhhH$, $\lhHH$) at LHC14.   The continuum and resonant production of CP-even Higgs boson pairs, $hh$ and $hH$, offer complementary probes of the scalar potential away from the light-Higgs decoupling limit.  We identify the viable search channels at LHC14 and estimate their expected discovery sensitivities.

\end{abstract}

\newpage

\section{Introduction}
\label{sec:intro}

Particle physics is at a crossroads.  The discovery of the 125.5 GeV Higgs boson at the LHC \cite{Aad:2012tfa,Chatrchyan:2012ufa} validates the fundamental theoretical tenet that the electroweak gauge symmetry is spontaneously broken.  The LHC measurements of the Higgs couplings to weak bosons, photons, gluons and fermions are all consistent with their Standard Model (SM) predicted values \cite{ATL14009,CMS:yva,Aad:2013wqa}.  This is satisfying, to a degree, but it is also mystifying that the SM should work so well. The Higgs mass is not predicted in the SM, and the large hierarchy of the electroweak and Planck scales is unexplained by the SM.  The commonly expected explanation for the hierarchy -- physics beyond the SM at the TeV scale -- has not been borne out thus far by the LHC experiments at 7 and 8 TeV cm energy. 
The LHC upgrade to 14 TeV (LHC14), with 10 times the present luminosity, may change this situation by the discovery of new particles.  Regardless, the properties of the Higgs boson will be central in the search for new physics (for a recent overview, see~\cite{Dawson:2013bba}).  The Higgs potential itself has so far not been subject to experimental scrutiny, since this requires the more challenging measurements of triple and quartic Higgs self-interactions via pair production of Higgs bosons.   However, this important avenue of pursuit should soon be possible with data from the upcoming LHC14 run.

The Two Higgs Doublet Model (2HDM) (see e.g.~\cite{higgshunters,Branco:2011iw,Gunion:2002zf}) provides a convenient general framework in which to explore extensions of the SM and to characterize deviations of the Higgs couplings from their SM values in analyses of experimental data.  For the case of a CP-conserving Higgs potential, the three physical neutral Higgs states consist of 2 CP-even states, $h$ and $H$, and a CP-odd state $A$. The pair-production of these Higgs bosons is the means by which the Higgs potential can be experimentally determined and signs of new physics may be found~\cite{Baur:2002rb,Baur:2002qd,Baur:2003gpa,Baur:2003gp}.  The goal of our study to is assess in which final states the Higgs pair production processes, especially $hh$ and $hH$, can be measured within a specific class of 2HDMs, the Type-II 2HDM (see e.g.~\cite{Lee:1973iz,Fayet:1974fj,Peccei:1977hh,Fayet:1976cr,Carena:2002es}), for which one of the Higgs doublets has tree-level couplings only to up-type quarks and the other has tree-level couplings only to down-type quarks and leptons (and thus it includes the minimal supersymmetric standard model as a special case).  Beyond this, a determination of the Higgs self-couplings can be made to some degree~\cite{Baur:2003gp,Contino:2012xk,Chen:2014xra}.  Implicit in this strategy is that the generalization of the Higgs sector is the only modification of new physics signals of relevance to Higgs pair production and decay; we will not consider scenarios where new physics in other sectors affects the production of Higgs pairs, as found in~\cite{Dib:2005re,Wang:2007zx,Wang:2007nf,Han:2009zp,Gillioz:2012se,Grober:2010yv,Asakawa:2010xj,Dawson:2007wh,Dawson:2012mk,Kribs:2012kz,Dolan:2012rv,Dolan:2012ac,Liu:2013woa,Cao:2013si,Ellwanger:2013ova,Han:2013sga}.

\begin{figure}[htbp]
\centering
\includegraphics[scale=0.25]{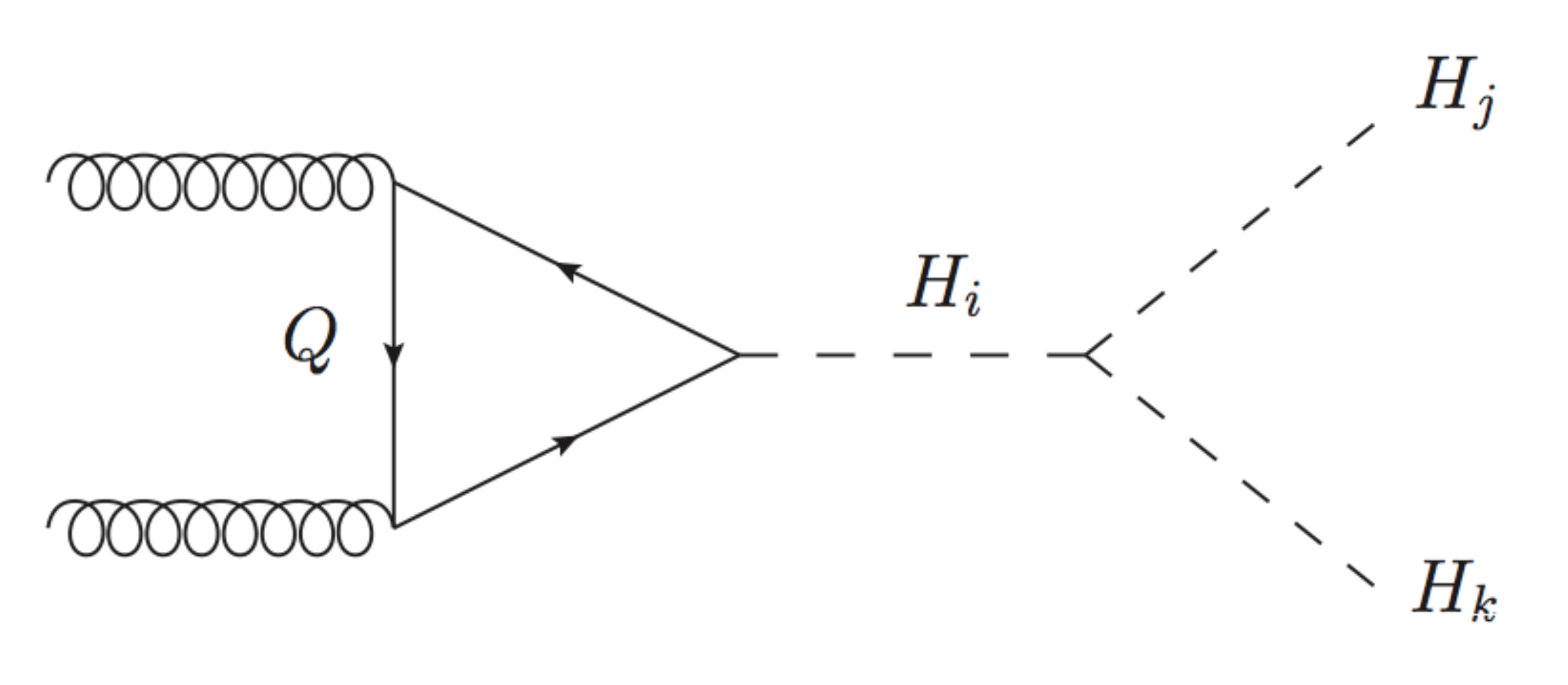}
\includegraphics[scale=0.25]{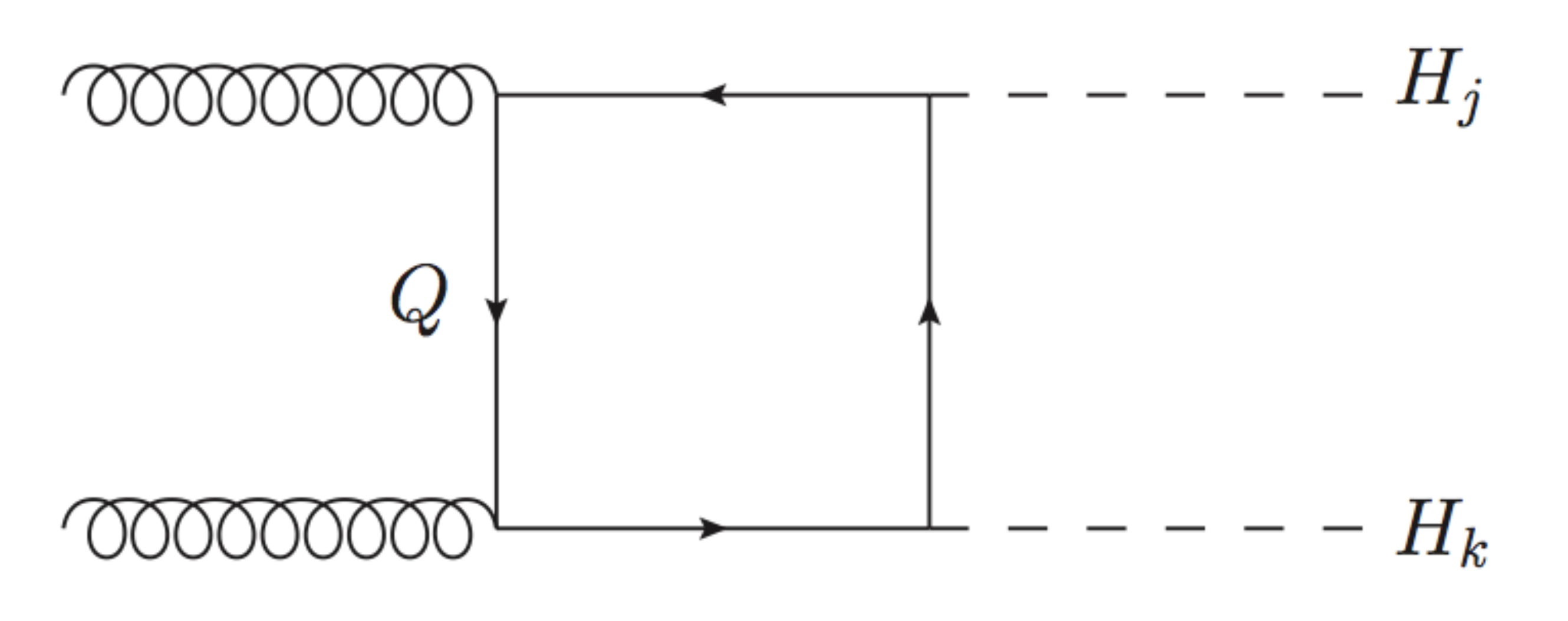}
\caption{Representative Feynman diagrams which contribute to Higgs boson pair production.}
\label{fig:feynman-diags}
\end{figure}

 The layout of our study is as follows.
In Section~\ref{sec:model}, we briefly describe the 2HDM, discuss the present constraints relevant for our study, and introduce three benchmark points to help elucidate discovery prospects.  
In Section~\ref{sec:xs}, we review and present the analytic formulae for the pair production of the Higgs bosons in gluon-gluon fusion, which is the dominant sub-process in $pp$ collisions at the LHC.  There are two classes of contributing Feynman diagrams: $s$-channel Higgs boson exchange and a box diagram with a top-quark loop, as shown in Fig.~\ref{fig:feynman-diags}. Representative Higgs pair production cross-sections are provided in this section.  
In Section~\ref{sec:hhsim}, we  describe our simulation of the $hh$ subprocess and subsequent $h$ decays.  We then proceed with a systematic consideration of the possible decay channels of the $hh$ along with their backgrounds from the relevant SM processes.  We then describe the Multi-Variate Analysis (MVA) methodology that is the basis of our extraction of the signal from the background.  The MVA methodology distinguishes signal from background by kinematics, and takes multiple variables into account simultaneously.

In Section~\ref{sect:hH}, we turn to a study of associated $hH$ production, which we find to be complementary to the resonant $H \to hh$ production.  In the $hH$ process, both the triangle and box diagrams in Fig.~\ref{fig:feynman-diags} contribute, with contributions of both $h^*$ and $H^*$ in the s-channel.  The triangle diagrams provide sensitivity to the products of the top-Yukawa and the $\lhhH$ tri-scalar couplings.    Depending on its mass, the heavy scalar, $H$, has several available decay channels that can potential provide identifiable signals, including $b\bar b, WW^*, ZZ^*$, and $t\bar t$. The decay branching fractions of $H$ to these channels are dependent on the mass and Higgs mixing parameters.  We perform simulations of these channels and their SM backgrounds to assess the discovery prospects.  We find that the following channels all lead to a possible discovery: $hH \to \bbaa, \bb\bb, ZZ\bb, \bb\tt$ and $ \bb\bbaa$, allowing for a rich variety of measurements.  Finally, in Section~\ref{sec:conclusion}, we summarize our results.  In our evaluation of the reach of LHC14, we assume throughout an integrated luminosity of 3 ab$^{-1}$.

\section{The Two Higgs Doublet Model}
\label{sec:model}

In this section, we will provide a very brief overview of the 2HDM and the theoretical constraints on the potential (for more comprehensive discussions, see e.g.~\cite{higgshunters,Branco:2011iw,Gunion:2002zf}).  The model consists of two Higgs doublets, which we express as the opposite-hypercharge Higgs doublets $\Phi_{1,2}$ as follows:
\be
\Phi_1 = \left( \begin{array}{c}
(\phi_1^{0} + v_{1} - i \eta_1^{0})/\sqrt{2}   \\ -\phi_1^-  \end{array} \right),\qquad 
\Phi_2 = \left( \begin{array}{c}
\phi_2^+ \\
(\phi_2^{0} + v_{2} + i \eta_2^{0})/\sqrt{2}     \end{array} \right),
\ee
in which the vacuum expectation values (vevs)  $v_{1,2}$ satisfy the relation $v = \sqrt{v_1^2+v_2^2} = 246$ GeV.  We follow standard practice and assume for simplicity both that CP is conserved ({\it i.e.}, is not explicitly or spontaneously broken), and that the theory obeys a softly broken $Z_2$ symmetry that eliminates quartic terms that are odd in either of the doublets, but allows a quadratic term that mixes $\Phi_1$ and $\Phi_2$ (this is consistent with our eventual specialization to the Type II 2HDM; see e.g.~\cite{Branco:2011iw} for a detailed discussion of these issues).  With these assumptions, the scalar potential takes the following form:
\bea
V &=&  m_1^2 \Phi_1^\dagger \Phi_1+m_2^2 \Phi_2^\dagger \Phi_2 - \frac{1}{2} M^2 \sin 2\beta (\Phi_1^\dagger \tilde{\Phi}_2 + \tilde{\Phi}_2^\dagger \Phi_1)
 +
 {\lambda_1 \over 2} |\Phi_1^\dagger \Phi_1|^2+{\lambda_2 \over 2} |\Phi_2^\dagger \Phi_2|^2
\nn
 \\ &+& 
 \lambda_3  |\Phi_1^\dagger \Phi_1 \Phi_2^\dagger \Phi_2| 
 +\lambda_4 | \Phi_1^\dagger \tilde{\Phi_2} \tilde{\Phi}_2^\dagger \Phi_1 |
 + {\lambda_5 \over 2} \left[(\Phi_1^\dagger \tilde{\Phi}_2)^2 + (\tilde{\Phi}_2^\dagger \Phi_1)^2\right],
\label{pot}
\eea
in which 
\bea
\tilde{\Phi}=i\sigma_2 \Phi^*.
\eea
After incorporating the minimization conditions, the scalar potential parameters can be replaced by physical masses and mixing angles.  There are two mixing angles: the angle $\beta=\tan^{-1}v_2/v_1$, and the angle $\alpha$, which is the mixing angle of the CP-even Higgs sector.  The quantity $\cba$ is of particular interest in that when $\cba\rightarrow 0$, the lightest neutral CP-even Higgs boson $h$ behaves like the Higgs boson of the SM, and the additional Higgs bosons decouple (for a comprehensive analysis of the CP-conserving 2HDM in the decoupling limit, see~\cite{Gunion:2002zf}).

Returning to the replacement of the $\lambda_i$ by the physical masses and mixing angles, it is convenient to parametrize them in terms of $v$, 
the $Z_2$-breaking potential parameter $M$, the Higgs masses $M_h, M_H, M_{H^\pm}, M_A$, and the angles $\alpha$, and $\beta$, as follows:
\bea
\lambda_1 &=& {-M^2 \tan^2\beta + M_h^2 \sin^2\alpha \sec^2\beta+M_H^2 \cos^2\alpha \sec^2\beta \over v^2},\\
\lambda_2 &=& {-M^2 \cot^2 \beta + M_h^2 \cos^2\alpha \csc^2\beta+M_H^2 \sin^2\alpha \csc^2\beta \over v^2},\\
\lambda_3 &=& {-M^2  + {1\over 4} (M_H^2 - M_h^2) \sin2\alpha \csc 2 \beta+2 M_{H^\pm}^2  \over v^2},\\
\lambda_4 &=& {M^2+M_A^2-2 M_{H^\pm}^2 \over v^2},\\
\lambda_5 &=& {M^2 - M_A^2\over v^2}.
\eea
In our analysis, we require for simplicity that the heavy physical mass scales are all equivalent, i.e. $M_H = M_A = M_{H^\pm}$, which  serves to ease any tension that would exist with electroweak precision data that prefers a small mass splitting. We see from the above expressions that $\lambda_3+\lambda_4+\lambda_5$, which is what appears in the trilinear scalar couplings of $h$ and $H$ (see e.g.~\cite{Branco:2011iw} for details), is
\bea
\lambda_3+\lambda_4+\lambda_5={M^2+(M_H^2-M_h^2)\csc 2\beta \sin 2\alpha \over v^2}, 
\eea
and thus is unaffected by the assumption that the heavier Higgs particles are mass-degenerate.  

\begin{figure}[htbp]
\begin{center}
\includegraphics[width=0.49\textwidth]{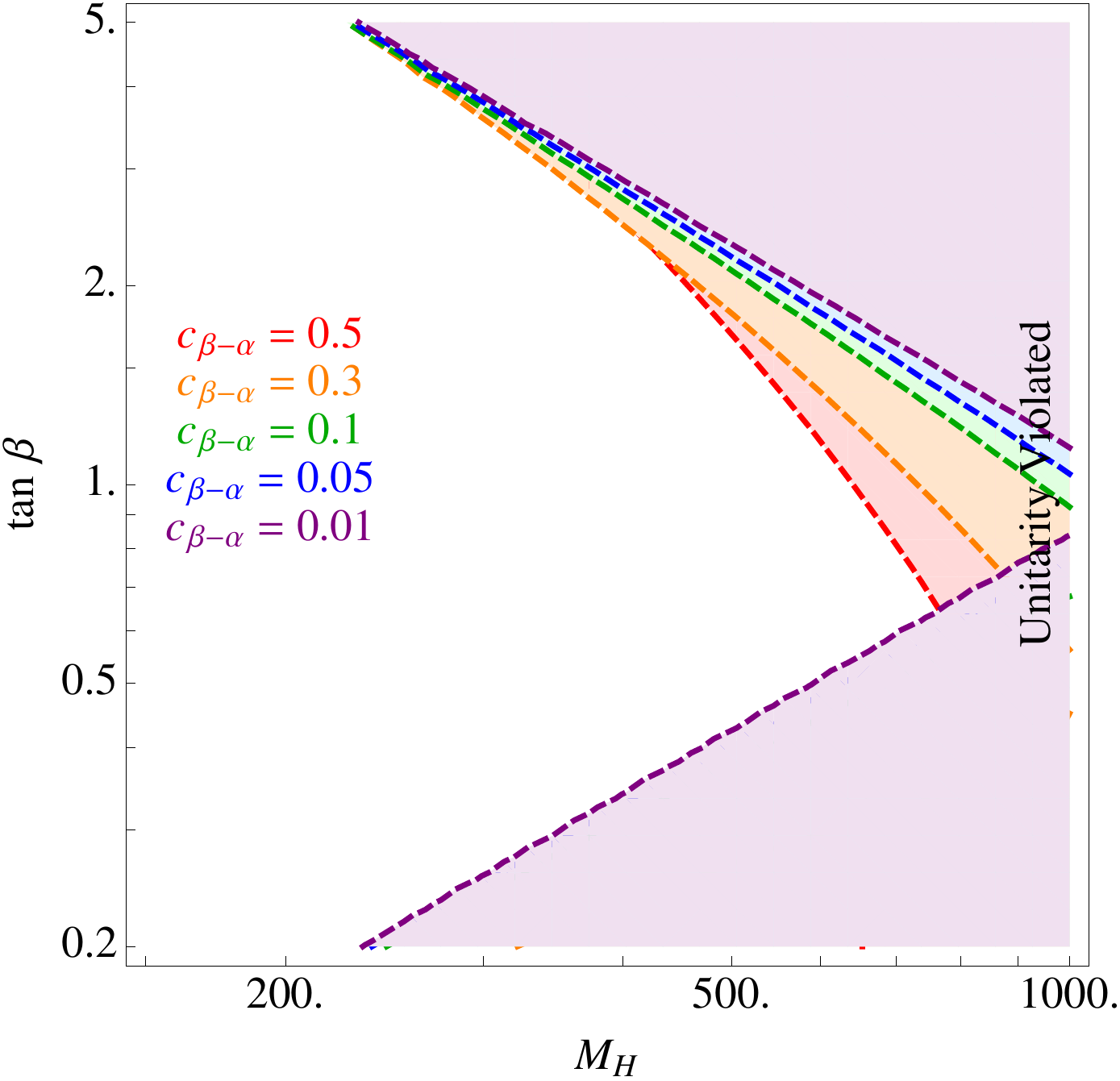}
\includegraphics[width=0.49\textwidth]{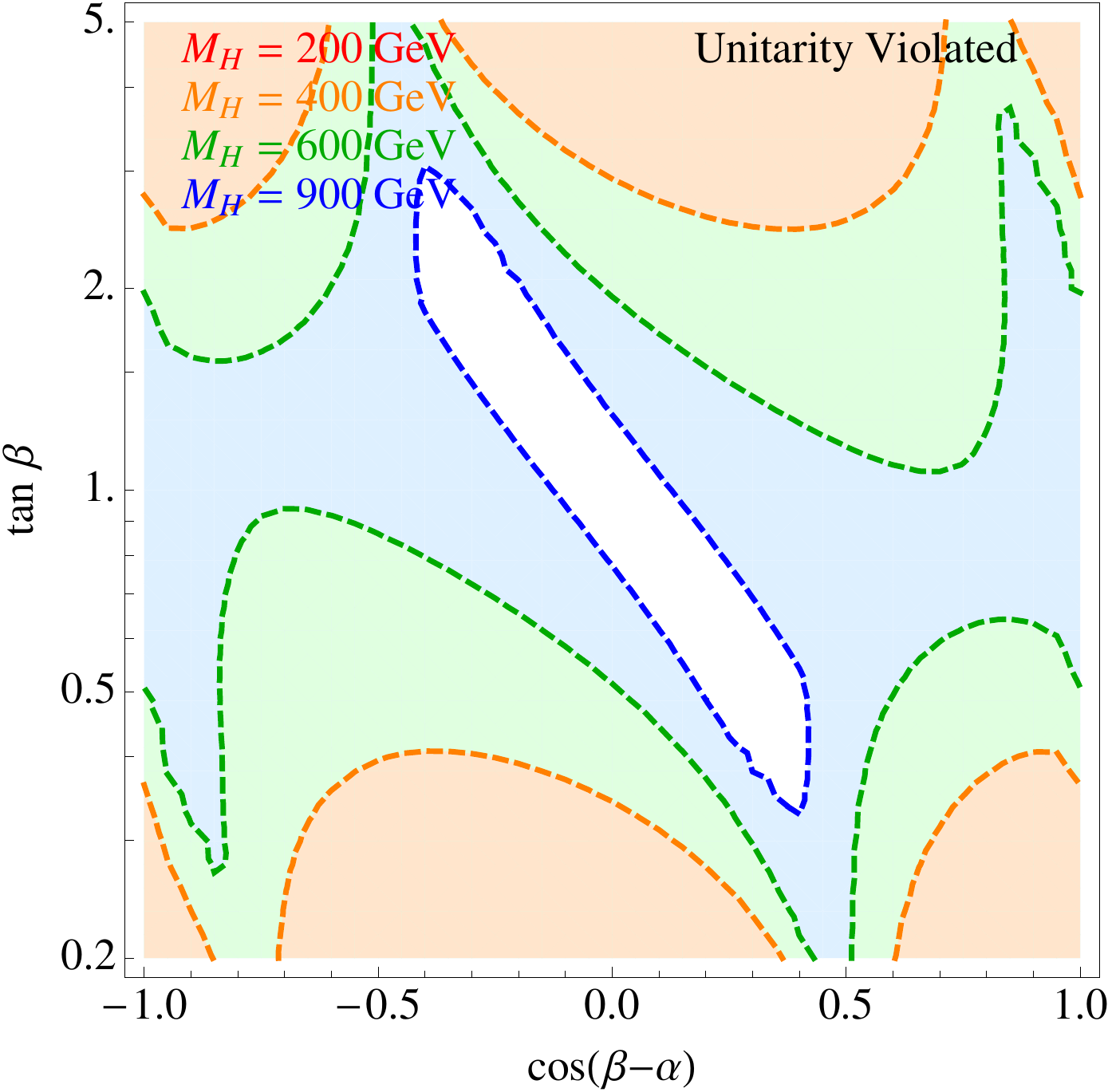}
\caption{Regions that violate perturbative unitarity are shaded in the colors corresponding to the listed values of $\cba$. The decoupling of the heavy Higgs sector for $\cba\rightarrow 0$ can allow for a rather large value of $M$. }
\label{fig:pertfig}
\end{center}
\end{figure}

We now impose the conditions that the potential maintains perturbative unitarity and is not unbounded from below.
As demonstrated in \cite{Arhrib:2000is} (and discussed in detail in~\cite{Branco:2011iw}), the conditions to be satisfied for perturbative unitarity are that the following quantities are $\leq 8\pi$:
\bea
a_\pm & = & \frac{3}{2} (\lambda_1+\lambda_2)\pm \frac{1}{2} \sqrt{9(\lambda_1-\lambda_2)^2+(2\lambda_3+\lambda_4)^2}\\
b_\pm & = &\frac{1}{2}\left ( \lambda_1+\lambda_2 \pm \sqrt{(\lambda_1-\lambda_2)^2+4\lambda_4^2} \right )\\
c_\pm & = & \frac{1}{2}\left ( \lambda_1+\lambda_2 \pm \sqrt{(\lambda_1-\lambda_2)^2+4\lambda_5^2} \right )\\
f_+ &=&  \lambda_3 +2\lambda_4+3\lambda_5, \quad f_- = \lambda_3 +\lambda_5, \quad f_{1}=f_2= \lambda_3 +\lambda_4,\\
e_{1}&=& \lambda_3+2\lambda_4-3\lambda_5, \quad e_{2}=2 \lambda_3-\lambda_5, \quad p_1 = \lambda_3-\lambda_4),
\eea
The necessary and sufficient conditions for the potential to remain unbounded from below are~\cite{Maniatis:2006fs}:
\bea
\lambda_1\ge 0,\quad \lambda_2 \ge 0&,&\quad \lambda_3 \ge -\sqrt{\lambda_1 \lambda_2},\\
\lambda_3+\lambda_4-|\lambda_5| &\ge& -\sqrt{\lambda_1\lambda_2}.
\eea
In Fig.~\ref{fig:pertfig}, we show the constraints arising from the requirement of perturbative unitarity in the $\tb$ v. $M_H$ and $\tb$ v.~$\cba$ planes for $M/M_H=0.8$.
We see that the $\tb-\cba$ plane is particularly instructive for inspecting Higgs couplings. Heavy state masses up to 1 TeV may be possible near $\tan \beta=1$.

\begin{figure}[htbp]
\begin{center}
\includegraphics[width=0.89\textwidth]{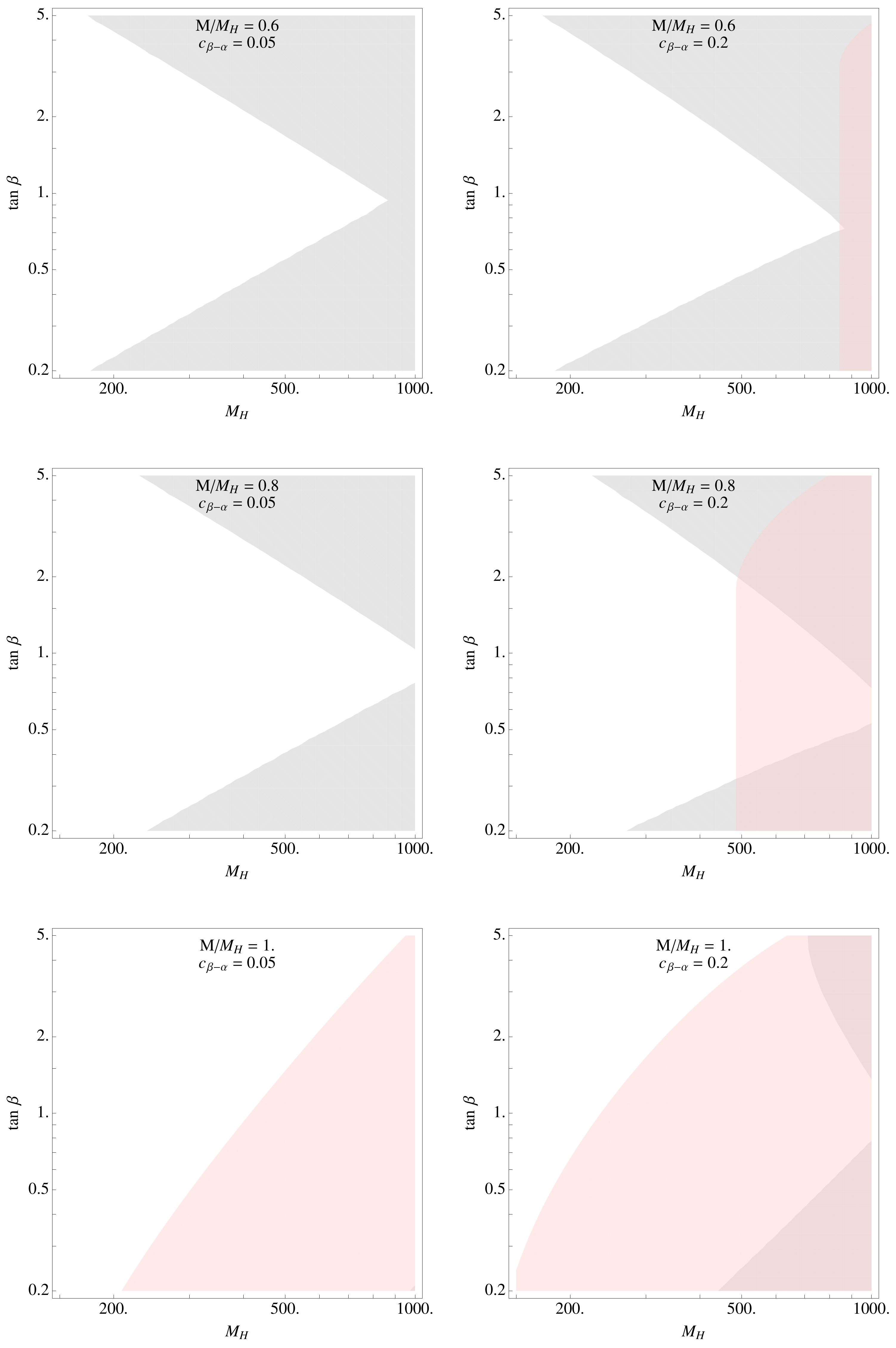}
\caption{Regions that violate perturbative unitarity (gray) and do not have a bounded potential (pink) are shaded for selected values of $\cba$ and $M/ M_H$.}
\label{fig:mhtbpertunit}
\end{center}
\end{figure}

In Fig.~\ref{fig:mhtbpertunit}, we demonstrate how both constraints combine to limit the available ranges of $\tb$ and $M_H$, for selected values of $\cba$ and $M/M_H$.   
For $M/M_H=1$, the bounded potential constraint severely limits the available parameter space, while the perturbative unitarity condition is substantially relaxed. For lower values of $M/M_H$, the potential constraint is not as severe.  For the remainder of this work, we fix $M/M_H = 0.8$ for illustrative purposes.

\subsection{Yukawa couplings}
\begin{figure}[htbp]
\begin{center}
\includegraphics[width=0.49\textwidth]{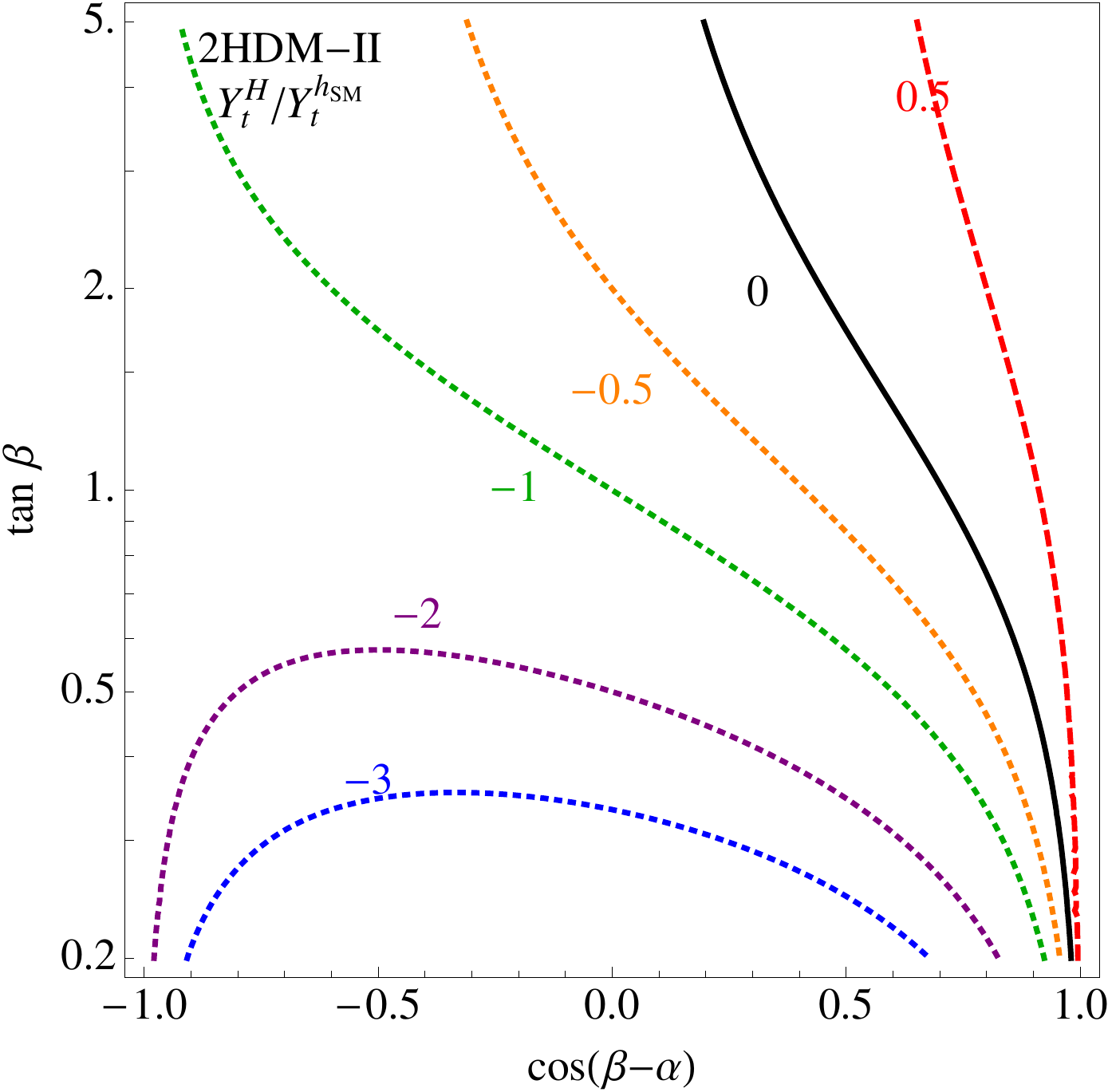}
\includegraphics[width=0.49\textwidth]{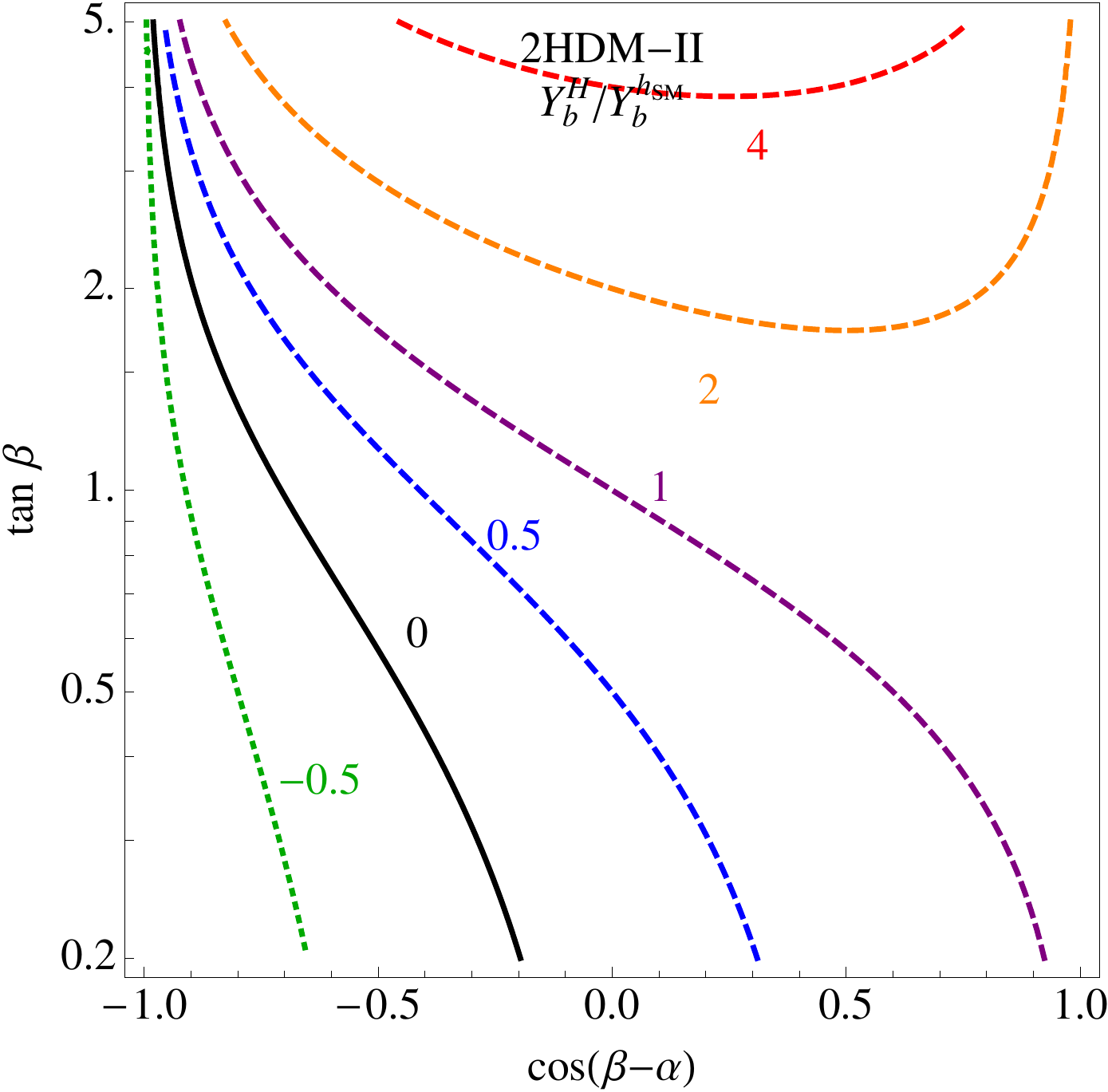}
\caption{Contours of the heavy Higgs Yukawa coupling to $t$ and $b$-quarks  in the plane of $\tan \beta$ and $M_H$ for selected values of $\cos(\beta-\alpha)$.  }
\label{fig:ytyb}
\end{center}
\end{figure}
For concreteness, we adopt the Yukawa sector of the Type-II 2HDM (the 2HDM-II), which takes the form (see e.g.~\cite{Lee:1973iz,Fayet:1974fj,Peccei:1977hh,Fayet:1976cr,Carena:2002es,Gunion:2002zf}):
\be
-{\cal L}_{\rm Yuk} = y_d \bar d_R \Phi_1 Q_L - y_u \bar u_R \Phi_2 Q_L + y_\ell \bar \ell_R \Phi_1 L_L.
\ee 
In Fig.~\ref{fig:ytyb}, we show the Yukawa couplings of the heavy Higgs boson to $t$ and $b$ quarks in this scenario as contours in the $\tb$ v.~$\cba$ plane.  

Within the 2HDM-II, measurements of the $h$ boson couplings with LHC Run-I data constrain the available ranges of $\cba$ and $\tb$.  A number of recent studies have found that the range of $\cba$ to be no more than 0.1 at 95\% C.L~in light of these data~\cite{Barger:2013ofa,Chen:2013rba,Craig:2013hca,Cheung:2013rva,Dumont:2014wha}. However, some additional freedom can be given in other versions of the 2HDM, such as the Type-I~\cite{Haber:1978jt} or Lepton specific models~\cite{Barnett:1983mm,Barnett:1984zy,Grossman:1994jb}.  A generic Yukawa aligned model with suppressed tree-level FCNCs is also consistent with the LHC data~\cite{Pich:2009sp,Cree:2011uy,Altmannshofer:2012ar,Bai:2012ex}.  

Complementarity of the gauge couplings forces a limit on the value of $\cba$ from the vector boson couplings of $h$ alone. We find that the combined ATLAS and CMS Run-I data~\cite{ATL14009} from vector boson coupling measurements provides a lower limit of $\kappa_V = \sin(\beta-\alpha) > 0.89$ at the 95\% C.L., which translates to an upper limit of
\be
|\cba| \lesssim 0.45.
\ee
In the 2HDM illustrations provided, these facts should be kept in mind for the larger values of $\cba$. For the $h$ state, for simplicity, we assume branching fractions consistent with the SM Higgs boson.  

\subsection{Scalar couplings}
\label{sect:scalarcoup}
The triscalar coupling, $\lambda^{hhh}$ in the SM takes the value
\be
\lambda^{hhh}_{\rm SM} = {3 M_h^2 \over v^2}.
\ee
Recent analyses of measuring this coupling at the LHC via the $hh$ continuum have shown that it may be possible to measure it with an uncertainty of order 30-50\%~\cite{Yao:2013ika,Barger:2013jfa}.  Substantial deviations away from the SM value allow a better determination due to interference effects~\cite{Barger:2013jfa}.  

In the 2HDM, this coupling is altered to
\bea
\lambda^{hhh} = 
 {3 M_h^2 \over 2v}\csc 2\beta (\cos(3\alpha-\beta)+3\cos(\alpha+\beta) )- {6 M^2\over v} \csc 2\beta \cos^2(\beta-\alpha) \cos(\alpha+\beta).
\label{eq:lhhhfull}
\eea
Expanding in the decoupling limit parameter $\cba\rightarrow 0$, the deviation of this coupling from its SM value is a second order effect.  It can be cast into the form
\bea
\lambda^{hhh}\approx {3 M_h^2 \over v^2} + \cos^2(\beta-\alpha) { 9 M_h^2 -12 M^2 \over 2 v}
\approx \lambda^{hhh}_{\rm SM} \left[1 + \cos^2(\beta-\alpha)  \left({3\over 2} - {2 M^2\over  M_h^2 }\right)\right],
\label{eq:lhhh}
\eea
in which higher order terms in $\cos(\beta-\alpha)$ have been dropped.  

\begin{figure}[htbp]
\begin{center}
\includegraphics[width=0.99\textwidth]{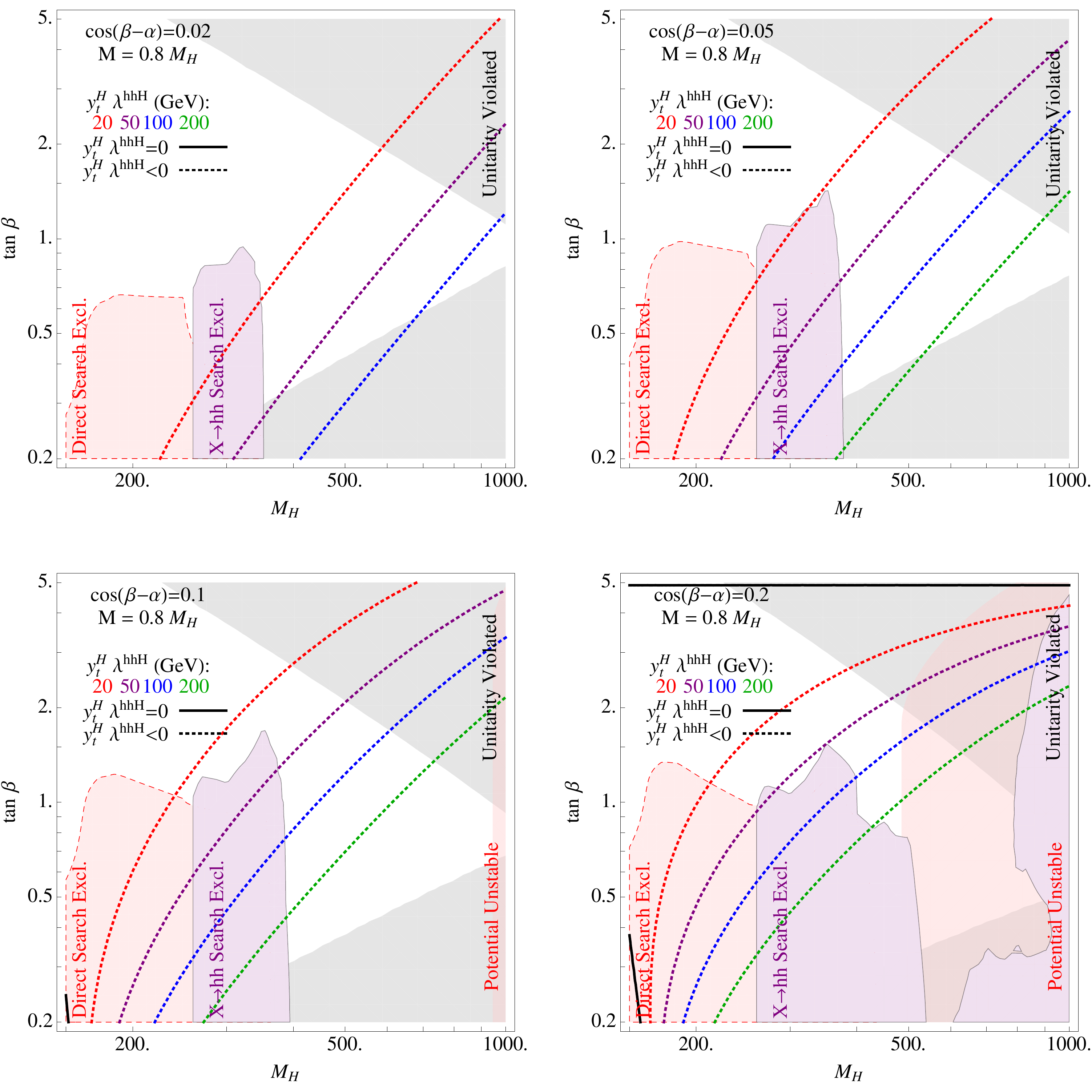}
\caption{Contours of $\lambda^{hhH}$ in the plane of $\tan \beta$ and $M_H$ (in GeV) for selected values of the decoupling parameter $\cos(\beta-\alpha)$.  Included are the unitarity (gray) and vacuum stability (pink) constraints assuming  $M=0.8 M_H$,  the direct search exclusion limits (dashed pink) from CMS~\cite{Chatrchyan:2013yoa} and the $hh\to b\bar b \gamma\gamma$ resonance search (purple)~\cite{CMS:2013eua,CMS:2014ipa,Aad:2014yja}.}
\label{fig:lhhH}
\end{center}
\end{figure}
The combination $y_t \lambda^{hhH}$ is the most relevant for the process of interest.  The possible values it may take are shown in Fig.~\ref{fig:lhhH} for selected values of $\cba = 0.02, 0.05, 0.1$ and 0.2.  We also show the excluded regions from the direct search of $H$ at the LHC~\cite{Chatrchyan:2013yoa} via vector boson decays, and from the search for a resonance in the $hh\to b\bar b \gamma\gamma$ final state~\cite{CMS:2013eua,CMS:2014ipa,Aad:2014yja}.

The  scalar couplings involving the heavy CP-even neutral Higgs that are important for additional search channels are given by
\bea
\label{eq:lhhH}
\lhhH&=&  {\cba\over\sin2\beta} \left ({M^2 (\sin2\beta-3\sin2\alpha)+(2 M_h^2+M_H^2)\sin2\alpha\over v} \right),\\
\label{eq:lhHH}
\lhHH&=&  {\sba\over\sin2\beta}\left ({M^2 (\sin2\beta+3\sin2\alpha)-(M_h^2+2M_H^2)\sin2\alpha\over v}\right ).
\eea
As previously discussed, these couplings have no $M_A$ and $M_{H^\pm}$ dependence (as they depend on the combination $\lambda_3+\lambda_4+\lambda_5$), and hence our assumption of heavy Higgs mass degeneracy does not affect these couplings.
\begin{figure}[htbp]
\begin{center}
\includegraphics[width=0.99\textwidth]{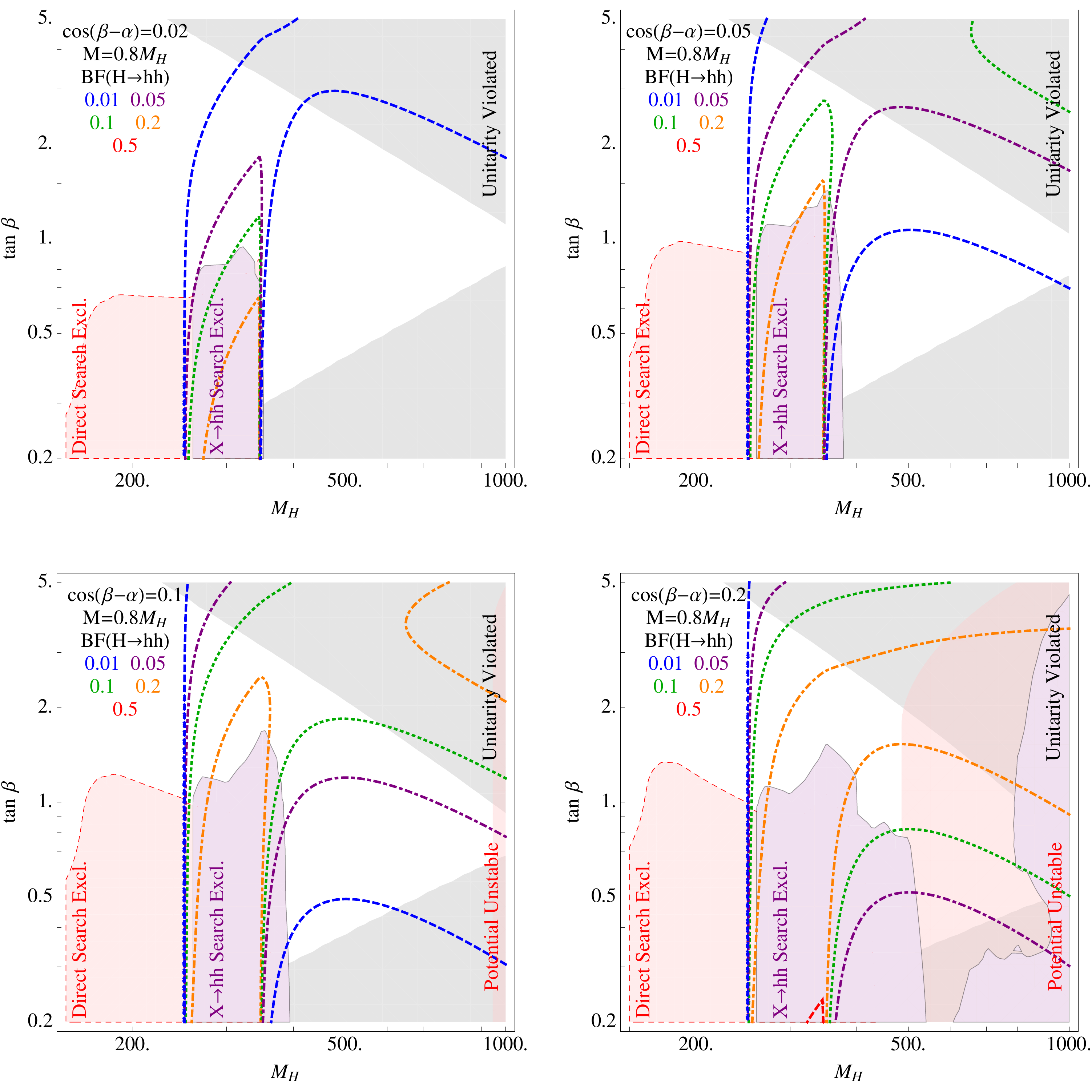}
\caption{Contours of $BF(H\to hh)$ in the plane of $\tan \beta$ and $M_H$ for selected values of $\cos(\beta-\alpha)$. Additional experimental and theoretical constraints are shown as in Fig.~\ref{fig:lhhH}.}
\label{fig:H2hh}
\end{center}
\end{figure}
In the decoupling limit, these expressions take the form
\bea
\lhhH&\approx& \cba{4 M^2-2 M_h^2-M_H^2\over v},\\
\lhHH&\approx &{-2 M^2+M_h^2+2M_H^2\over v} + \cba {2(-3M^2+M_h^2+2M_H^2)\cot 2\beta \over v},
\label{eq:lhHH2}
\eea
neglecting terms of $O(\cba^2)$.  
Hence, near the decoupling limit the $hhH$ coupling is suppressed while the $hHH$ coupling persists~(see e.g.~\cite{Gunion:2002zf,Carena:2013ooa} for discussions).  This is shown in Fig.~\ref{fig:H2hh}, which gives the contours of ${\rm BF}(H\to hh)$; additional details of the $H$ decay modes are discussed in Appendix~\ref{apx:bfs}.  
The window of $2 m_h < M_H < 2 m_t$ in which the ${\rm BF}(H\to hh)$ is quite large and in some cases already ruled out for low $\tb$.  In Section~\ref{sect:bbaa}, we will see that the discovery potential roughly follows this region, but with a few caveats.

By extracting the $\lhhh, \lhhH$ and $\lhHH$ couplings to some degree of precision, the self-consistency of the scalar model may be tested.   More precisely, by measuring the physical masses $M_h$ and $M_H$ and the heavy Higgs coupling to vector bosons, it is possible to determine whether the expressions given in Eqs.~\ref{eq:lhhh}, \ref{eq:lhhH} and \ref{eq:lhHH} are self-consistent.  

In the subsequent analyses, we will refer to three benchmark points that help elucidate the discovery potential of each channel. The points are summarized in Table~\ref{tab:bench}.  Benchmark point A will illustrate the viability of the $H\to hh\to \bbaa$ channel, point B the $hh/hH\to \bbaa$ and $b\bar b b\bar b$ channels, and point C the $hH\to t\bar t b\bar b$ channel.

\begin{table}[h]
\renewcommand{\arraystretch}{1.33}
\begin{center}
\begin{tabular}{|c |c |c| c|}
\cline{2-4}
\multicolumn{1}{ c| }{} & {\bf A}:&{\bf B}:&{\bf C}: \\[-1.5mm]
 \multicolumn{1}{ c| }{} & $M_H = 300$ GeV,&$M_H = 300$ GeV,& $M_H = 500$ GeV, \\[-2mm]
\multicolumn{1}{ c| }{} &$t_\beta = 2$, $c_{\beta -\alpha}= 0.1$&$t_\beta = 1$, $c_{\beta -\alpha} = 0.02$& $t_\beta = 1$, $c_{\beta -\alpha}= 0.02$\\
\hline
 $\lhhh / \lhhh_{SM}$  &0.946&0.998&0.992 \\
$\lhhH$ (GeV) &40.8&8.87&29.2 \\
$\lhHH$ (GeV) &310&327&795\\
$y^H_t$ &$-0.40$&$-0.98$&$-0.98$\\
   \hline
$\sigma(pp \to hh)$ (fb)& 340 &810 & 37 \\
$\sigma(pp\to hH)$ (fb)& 7.7 & 44 & 26 \\
\hline
$BF(H\to hh)$&18\%&7.6\%&0.1\% \\
$BF(H\to tt)$&0.0\%&0.0\%&99\% \\
$BF(H\to bb)$&34\%&74\%&0.2\% \\
$BF(H\to ZZ+WW)$&49\%&18\%&0.2\% \\
\hline
\end{tabular}
\end{center}
\caption{Benchmark points of relevant couplings, production cross sections at LHC14 and branching fractions for the channels of interest.}
\label{tab:bench}
\end{table}%

\section{Higgs Pair Production Cross Section}
\label{sec:xs}
Pairs of neutral Higgs bosons can be generated through two different loop processes (depicted in Fig.~\ref{fig:feynman-diags}): ({\it i}) the triangle diagram where an $s$-channel Higgs boson decays into two Higgs bosons and ({\it ii}) the box diagram where annihilation of two gluons through a square loop produces a Higgs boson pair.  The exact expressions for these one-loop diagrams with generic internal/external Higgs bosons (as well as generic heavy quarks) were first computed in Ref.~\cite{Plehn:1996wb}.  We have independently confirmed the expressions for the loop diagrams and we present them here just for completeness.  Readers interested in the finer details are referred to Sections 3 and 4 of Ref.~\cite{Plehn:1996wb}.

First, let us introduce some notation.  Denoting the intial-state gluon momenta as $p_{a,b}$ and the final-state Higgs boson momenta as $p_{j,k}$, the Mandelstam invariants are given by:
\bea
\hat{s} = (p_a + p_b)^2 \,\,\, ; \,\,\, \hat{t} = (p_a - p_j)^2 \,\,\, ; \,\,\, \hat{u} = (p_a - p_k)^2 \,. \nonumber
\eea
It is also useful to define the quantities
\bea
S = \hat{s}/m_Q^2 \,\,\, ; \,\,\, T = \hat{t}/m_Q^2 \,\,\, ; \,\,\, U = \hat{u}/m_Q^2 \,, \nonumber
\eea
\bea 
\rho_j = M_j^2 \,\,\, ; \,\,\, \rho_k = M_k^2 \,\,\, ; \,\,\, \tau_Q = 4/S \,, \nonumber
\eea
\bea
T_1 = T - \rho_j \,\, , \,\, U_1 = U - \rho_j \,\, , \,\, T_2 = T - \rho_k \,\, , \,\, U_2 = U - \rho_k. \nonumber
\eea
In the discussion to follow, we will reduce all tensor integrals to scalar ones.  The pertinent three- and four-point scalar integrals can be written as
\bea
C_{\ell m} &=& \int \frac{d^4 k}{(2\pi)^4} \frac{1}{(k^2 - m_Q^2) \left( (k + p_\ell)^2 - m_Q^2 \right) \left( (k + p_\ell + p_m)^2 - m_Q^2 \right)} \,, \nonumber \\
D_{\ell m n} &=&\int \frac{d^4 k}{(2\pi)^4} \frac{1}{(k^2 - m_Q^2) \left( (k + p_\ell)^2 - m_Q^2 \right) \left( (k + p_\ell + p_m)^2 - m_Q^2 \right) \left( (k + p_\ell + p_m + p_n)^2 - m_Q^2 \right)} \,, \nonumber
\eea
in which $\ell, m, n$ label momenta entering the loop.

The matrix element of the triangle diagram can be written in terms of a ``coupling'' $C_\triangle$ and a form factor $F_\triangle$ as:
\bea
{\cal M}_\triangle = \frac{G_F \alpha_s \hat{s}}{2 \sqrt{2} \pi} C_\triangle F_\triangle A_{1,\mu \nu} \epsilon_a^\mu \epsilon_b^\nu \delta_{ab} \, ,
\eea
in which the tensor structure $A_1^{\mu\nu}$ is:
\bea
A_1^{\mu\nu} = g^{\mu\nu} - \frac{p_a^\nu p_b^\mu}{(p_a \cdot p_b)} \,.
\label{eq:A1}
\eea
The coupling factor can be expressed as:
\bea
C_\triangle = \sum_{i = h,H} C^i_\triangle \,,
\eea
with:
\bea
C^i_\triangle = \lambda_{H_i H_j H_k} \frac{M_Z^2}{\hat{s} - M_{H_i}^2 + i M_i \Gamma_{H_i}} y_{H_i Q \bar{Q}} \,,
\eea
in which $y_{H_i Q \bar{Q}}$ denotes the heavy quark Yukawa coupling to $H_i$.  The form factor $F_\triangle$ can be computed in closed form, and is given by
\bea
F_\triangle = \frac{2}{S} \left[ 2 + (4 - S) m_Q^2 C_{jk} \right] = \tau_Q \left[ 1 + (1 - \tau_Q) f(\tau_Q) \right] \,, 
\eea
in which
\bea
f(\tau_Q) = \left\{
\begin{array}{cc}
\arcsin^2 \frac{1}{\sqrt{\tau_Q}} \,\,\,\,\, & \,\,\,\,\,\, \tau_Q \ge 1 \\
- \frac{1}{4} \left[ \log \frac{1 + \sqrt{1 - \tau_Q}}{1 - \sqrt{1 - \tau_Q}} - i \pi \right]^2  \,\,\,\,\, & \,\,\,\,\,\, \tau_Q < 1  \, .
\end{array}
\right.
\eea

The matrix element for the box diagrams can be written in terms of a coupling factor $C_\square$ and two gauge-invariant form factors $F_\square$ and $G_\square$ as:
\bea 
{\cal M}_\square =  \frac{G_F \alpha_s \hat{s}}{2 \sqrt{2} \pi} C_\square \left( F_\square A_{1,\mu\nu} + G_\square A_{2,\mu\nu} \right) \epsilon_a^\mu \epsilon_b^\nu \delta_{ab} \, ,
\eea
where $A_{1,\mu\nu}$ is given in Eq.~(\ref{eq:A1}) and the other tensor structure takes the form
\bea
A_2^{\mu\nu} = g^{\mu\nu} + \frac{p_j^2 p_a^\nu p_b^\mu}{p_T^ (p_a \cdot p_b)} - \frac{2 (p_b \cdot p_j) p_a^\nu p_j^\mu}{p_T^ (p_a \cdot p_b)} - \frac{2 (p_a \cdot p_j) p_b^\mu p_j^\nu}{p_T^ (p_a \cdot p_b)} + \frac{2 p_j^\mu p_j^\nu}{p_T^2} \,,
\eea
with
\bea
p_T^2 = 2 \frac{(p_a \cdot p_j) ( p_b \cdot p_j )}{(p_a \cdot p_b)} - p_c^2 \,.
\eea
The advantage of writing the amplitude in terms of $A_{1,2}^{\mu\nu}$ is that it greatly simplifies the calculation of the matrix-element-squared, since
\bea
A_1 \cdot A_2 = 0 \,\,\,\,\, {\rm and} \,\,\,\,\, A_1 \cdot A_1 = A_2 \cdot A_2 = 2 \,.
\eea

The coupling for the box diagrams is just the product of the two Yukawa couplings of the heavy quark to the two Higgs bosons
\bea
C_\square =   y_{H_j Q \bar{Q}} \, y_{H_k Q \bar{Q}}  \,,
\eea
while the form factors $F_\square$ and $G_\square$ are given by
\bea
F_\square &=& \frac{1}{S^2} \biggl\{ 4 S + 8 S m_Q^2 C_{ab} - 2 S m_Q^4 \left( S + \rho_j + \rho_k - 8 \right) \left( D_{abj} + D_{baj} + D_{ajb} \right) \nonumber\\
&+& \left( \rho_j + \rho_k - 8 \right) m_Q^2 \left[ T_1 C_{aj} + U_1 C_{bj} + U_2 C_{ak} + T_2 C_{bk} - m_Q^2 \left( TU - \rho_j \rho_k \right) D_{ajb} \right] \biggr\}  \,,
\eea
and
\bea
G_\square &=& \frac{1}{S(TU - \rho_j \rho_k)} \biggl\{ m_Q^2 \left( T^2 + \rho_j \rho_k - 8 T \right) \left[ S C_{ab} + T_1 C_{aj} + T_2 C_{bk} - S T m_Q^2 D_{baj} \right] \nonumber\\
&& + m_Q^2 \left( U^2 + \rho_j \rho_k - 8 U \right) \left[ S C_{ab} + U_1 C_{bj} + U_2 C_{ak} - S U m_Q^2 D_{abj} \right] \nonumber\\
&& - m_Q^2 \left( T^2 + U^2 - 2 \rho_j \rho_k \right) \left( T + U - 8 \right) C_{jk} \nonumber\\
&& - 2 m_Q^4 \left( T + U - 8 \right) \left( TU - \rho_j \rho_k \right) \left( D_{abj} + D_{baj} + D_{ajb} \right) \biggr\} \,.
\eea
The differential cross section (averaging/summing over initial/final state spins and colors) then takes the following form:
\bea
\frac{d\hat{\sigma}(gg \to H_j H_k)}{d\hat{t}} = \frac{G_F^2 \alpha_s^2}{256 (2\pi)^3} \biggl[ \bigg|  \left( C_\triangle F_\triangle + C_\square F_\square \right) \biggr|^2 + \bigg| C_\square G_\square \biggr|^2 \biggr] \,,
\eea
To obtain the total parton-level cross section, this expression is integrated over the scattering angle of one of the Higgs bosons.  Finally, to convert the parton-level cross section to the proton-proton cross section, we convolute the former with the PDFs for two gluons and integrate over the momentum fraction of the gluons.  For the parton distributions, we use CTEQ 6L1.

\section{Light Higgs pair production simulation}
\label{sec:hhsim}

\begin{figure}[htbp]
\begin{center}
\includegraphics[width=0.89\textwidth]{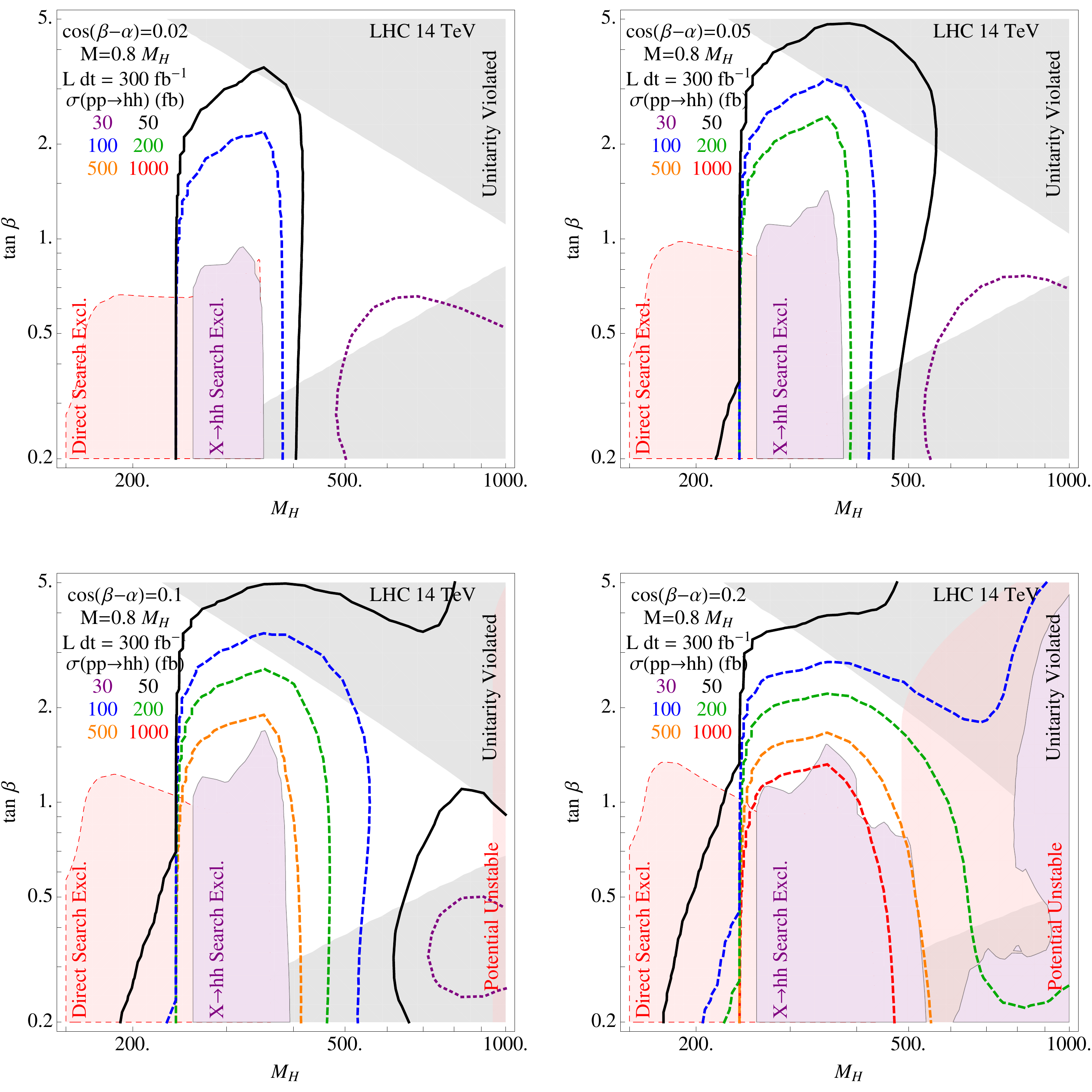}
\caption{Contours of $\sigma(pp\to  hh)$ in the plane of $\tan \beta$ and $M_H$ for selected values of $\cos(\beta-\alpha)$. Additional experimental and theoretical constraints are shown as in Fig.~\ref{fig:lhhH}. }
\label{fig:xscontour}
\end{center}
\end{figure}

The leading order (LO) matrix elements of the $hh$ subprocesses in Fig.~\ref{fig:feynman-diags} are known \cite{Plehn:1996wb,Eboli:1987dy,Glover:1987nx,Dicus:1987ic,Hespel:2014sla}.  We generate signal events by incorporating the loop amplitudes directly into MADGRAPH ~\cite{Alwall:2011uj}, and we include the NNLO K-factor of 2.27 for 14 TeV~\cite{Dawson:1998py,Dittmaier:2011ti,Branco:2011iw,Shao:2013bz,deFlorian:2013uza,Goertz:2013kp,deFlorian:2013jea}.  We note that in principle, the resonant production can shift the overall K-factor as the ratio $\sigma_{NNLO}/\sigma_{LO}$ can be $\sqrt{s}$ dependent.  However, since the K-factor has not been given for this process, we adopt the SM value and assume any shift induced by the $H$ resonance is small.  We show the cross section contours of $pp\to hh$ with the $H\to hh$ resonance in Fig.~\ref{fig:xscontour}.

The $pp\to hh$ cross section can be shifted dramatically away from its SM value by the presence of an extended Higgs sector \cite{Barger:2014taa,Moretti:2004wa,Arhrib:2009hc,Bhattacherjee:2014bca,Baglio:2014nea}. The relative competition of the diagrams in Fig.~\ref{fig:feynman-diags} strongly impacts the kinematic distributions with the most apparent coming from the resonant $gg\to H\to hh$ diagram.  Here, if $M_H > 2 m_h$, the resonance can become prominent, overwhelming the continuum from the $gg\to hh$ box and $gg\to h^*\to hh$ diagrams, seen as the large cross section in the $250-350$ GeV range, above which, the $H\to t\bar t$ branching fraction dominates.  

We note that the sign of the combination $y_t^H \lhhH$ determines the shape of the distribution due to the interference with the continuum diagrams.  In principle, measuring the $H\to hh$ lineshape can determine the sign of $y_t^H \lhhH$, further constraining the model.  A simple counting of events above and below resonance will provide a handle on the sign of the coupling combination, while more sophisticated fits including the matrix elements are possible, as has been done in the continuum case~\cite{Barger:2013jfa}. For sufficiently heavy $H$, the lower energy $M_{hh}$ distribution converges to the SM expectation.  We explore model independent resonant production of $hh$ in more detail in Ref.~\cite{Barger:2014taa}.  

Each final state Higgs boson in these events is decayed in the narrow width approximation to SM Higgs decay modes.  There are a number of potential final states for the Higgs pair, but most suffer suppression due to small SM branching fractions \cite{Liu:2013woa}. As noted in Ref.~\cite{Barger:2013jfa}, the $\bbll$ channel is swamped by the reducible background of $b\bar b jj$ where both light flavored jets fake a $\tau$.  While the fake rate is in the range of $1-3\%$, the total cross section of $b\bar b j j$ is at the $\mu b$ level. Moreover,  we neglect the $b\bar b W^+W^-$ channel due to a small SM significance~\cite{Baglio:2012np}.  The $\bb\bb$ channel also suffers from a large QCD background, and would only be viable with the use of jet substructure techniques~\cite{deLima:2014dta}. Therefore, we concentrate on the analysis of the $\bbaa$ channel for the resonant production of $hh$. Ref.~\cite{Chen:2013emb},  exploring the same channel, appeared while this work was in preparation.

\subsection{The $hh\to \bbaa$ channel}\label{sect:bbaa}
We simulate the pertinent backgrounds for the $\bbaa$ channel.  The irreducible background includes the following production modes:
\bea
pp&\to&\bbaa,\\
pp&\to& Z + h \to b\bar b +\gamma\gamma,
\eea
while the reducible backgrounds include
\bea
pp&\to& t\bar t + h ~\to~ b \ell^+ \nu ~\bar b \ell^- \bar \nu + \gamma \gamma \quad(\ell^\pm~ {\rm missed}),\\
pp&\to&b\bar b + j j ~\to~ b\bar b + \gamma \gamma\quad (j \to \gamma).
\eea
We assume a photon tagging rate of 85\% and a jet to photon fake rate of $\epsilon_{j\to \gamma}=1.2 \times 10^{-4}$~\cite{Aad:2009wy}.  We have determined the additional reducible backgrounds of $jj\gamma\gamma$ and $c\bar c \gamma\gamma$  are subdominant, therefore they are not included in this analysis.

To account for $b$ jet tagging efficiencies, we assume a $b$-tagging rate of 70\% for $b$-quarks with $p_T > 30~{\rm GeV}$ and $|\eta_{b}| < 2.4$ consistent with multivariate tagging suggested for the LHC luminosity upgrade~\cite{atlphyspub2013004}.  We also apply a mistagging rate for charm-quarks as
\begin{equation}
\epsilon_{c\to b} =10\% \quad\quad {\rm for } \quad p_T(c) > 50 {\rm GeV},
\end{equation}
while the mistagging rate for a light quark is: 
\begin{eqnarray}
\epsilon_{u,d,s,g\to b} &= 2\% \quad\quad \quad {\rm for }& p_T(j)> 250 {\rm GeV}\\
\epsilon_{u,d,s,g\to b} &= 0.67\% \quad\quad {\rm for }& p_T(j) < 100 {\rm GeV}.
\end{eqnarray}
Over the range $100~{\rm GeV}<p_T(j)<250~{\rm GeV}$, we linearly interpolate the fake rates given above~\cite{Baer:2007ya}. With pile-up the rejection rate is expected to worsen by up to 20\%~\cite{atlphyspub2013004}. Finally, we model detector resolution effects by smearing the final state energy according to
\begin{equation}
{\delta E \over E} = {a \over \sqrt{E}} \oplus b,
\end{equation}
where we take $a=50\%$ and $b=3\%$ for jets and $a=10\%$ and $b=0.7\%$ for photons.

We apply a multi-variate analysis (MVA) which relies on relevant kinematic variables.  We begin with low level cuts, requiring two $b$-tags and two $\gamma$-tags and no tagged charged leptons, with separation of $\Delta R_{\gamma\gamma}, \Delta R_{b\bar b}, \Delta R_{b\gamma} > 0.4$.  The value $\Delta R_{ab}=\sqrt{(\phi_a-\phi_b)^2+(\eta_a-\eta_b)^2}$ is the separation of two objects in the $\eta-\phi$ plane.  We further require $p_T(b,\gamma)> 30$ GeV and $|\eta_{b,\gamma}| < 2.4$.

We define a window within which the MVA will analyze events.  This window has the Higgs boson reconstructed in the $b\bar b$ and $\gamma\gamma$ channels according to:
\bea
|M_{b\bar b} - M_h| &<& 20~{\rm GeV},\\
|M_{\gamma\gamma} - M_h| &<& 10~{\rm GeV}.
\eea

We extend our analysis to include multiple variables simultaneously.  This allows one to in essence blend cuts together rather than perform a hard cut on a kinematic distribution.  We form a discriminant based on a set of observables which include: 
\be{\cal O}=\left\{ M_{\bbaa}, M_{b\bar b}, M_{\gamma\gamma}, p_T(b\bar b), p_T(\gamma\gamma), \Delta R_{b\bar b},\Delta R_{\gamma\gamma}, \Delta\eta_{\gamma\gamma},\Delta \eta_{b\bar b}\right\}.
\ee
The discriminant is then constructed by the ratio
\be
{\cal D}={S({\cal O})\over S({\cal O}) + A~ B({\cal O})},
\label{eqn:discrim}
\ee
in which $S({\cal O})$ and $B({\cal O})$ are the normalized differential cross sections in the observable space ${\cal O}$.  These differential cross sections are estimated via event generation.  The discriminator is evaluated for an event sample, yielding a value close to 1 for signal-like events and close to 0 for background-like events.  For the particular choice of $A=N_B / N_S$, the discriminant gives the probability of an event being signal~\cite{Barlow:1986ek}.  A cut may be placed on the value of ${\cal D}$, thereby selecting a relatively high signal event sample.  Such a multivariate discriminator can offer similar sensitivity that the matrix-element, or neural network methods allow~\cite{Abazov:2008kt}.

In practice, we apply a simplified version of the discriminant in which we ignore the correlations among the variables. With limited statistics, this allows a more efficient construction of the discriminator, defined as
\be
{\cal D}= {S\{{\cal O}_i\}\over S\{{\cal O}_i\} + B\{{\cal O}_i\}},
\ee
where $\{{\cal O}_i\}$ is the combinatorial subset of observables ${\cal O}$ that  go into the multivariate discriminant. In the MVA results that follow, further optimization may be done by including the correlations between observables, but we adopt this uncorrelated approach for simplicity.  We define the level of statistical significance, ${\cal S}$, according to~\cite{Bartsch:2005xxa}
\be
{\cal S} = 2 \left(\sqrt{S+B}-\sqrt{B}\right),
\ee
in which $S$ and $B$ are the number of signal and background events surviving cuts.  We maximize ${\cal S}$ by varying the cut on the discriminator, ${\cal D}_{\rm cut}$, which minimizes the choice of $A$ in Eq.~\ref{eqn:discrim}.

\begin{figure}[htbp]
\begin{center}
\includegraphics[width=0.99\textwidth]{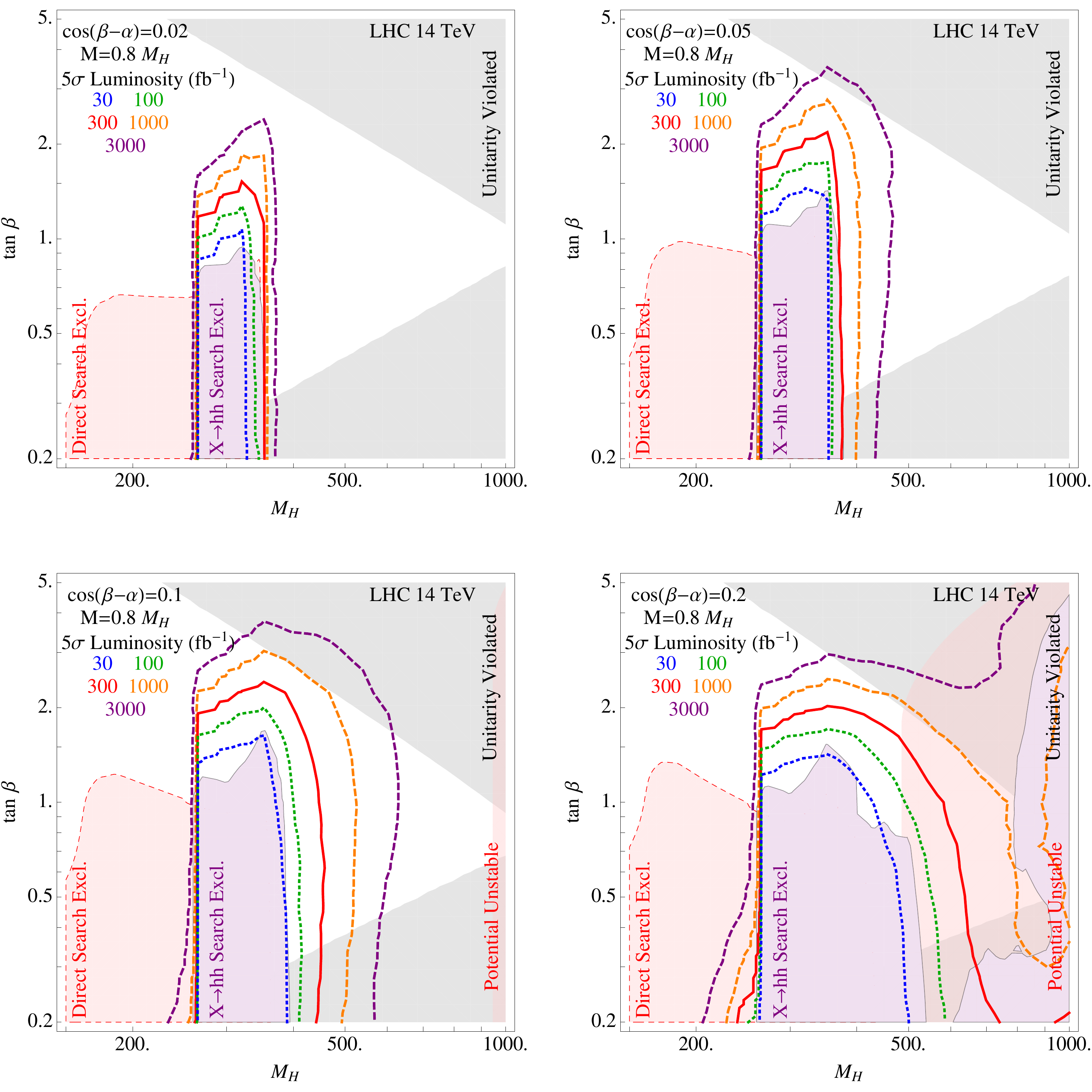}
\caption{Contours of the luminosity required for $5\sigma$ discovery in the plane of $\tan \beta$ and $M_H$ for selected values of $\cos(\beta-\alpha)$. Additional experimental and theoretical constraints are shown as in Fig.~\ref{fig:lhhH}.   }
\label{fig:lumin}
\end{center}
\end{figure}
In Fig.~\ref{fig:lumin}, we show the luminosity required to obtain $5\sigma$ discovery at the LHC.  We find that generally these contours follow the shape of the $hh\to b\bar b + \gamma\gamma$ resonance excluded region (shaded in purple) with Run-I data.  The contour with $\sqrt s = 14$ TeV and $30~{\rm fb}^{-1}$ of integrated luminosity is a close match with the 7+8 TeV exclusion region,  with small fluctuations likely caused by different analyses and statistical fluctuations in the data. 

\begin{figure}[htbp]
\begin{center}
\includegraphics[width=0.99\textwidth]{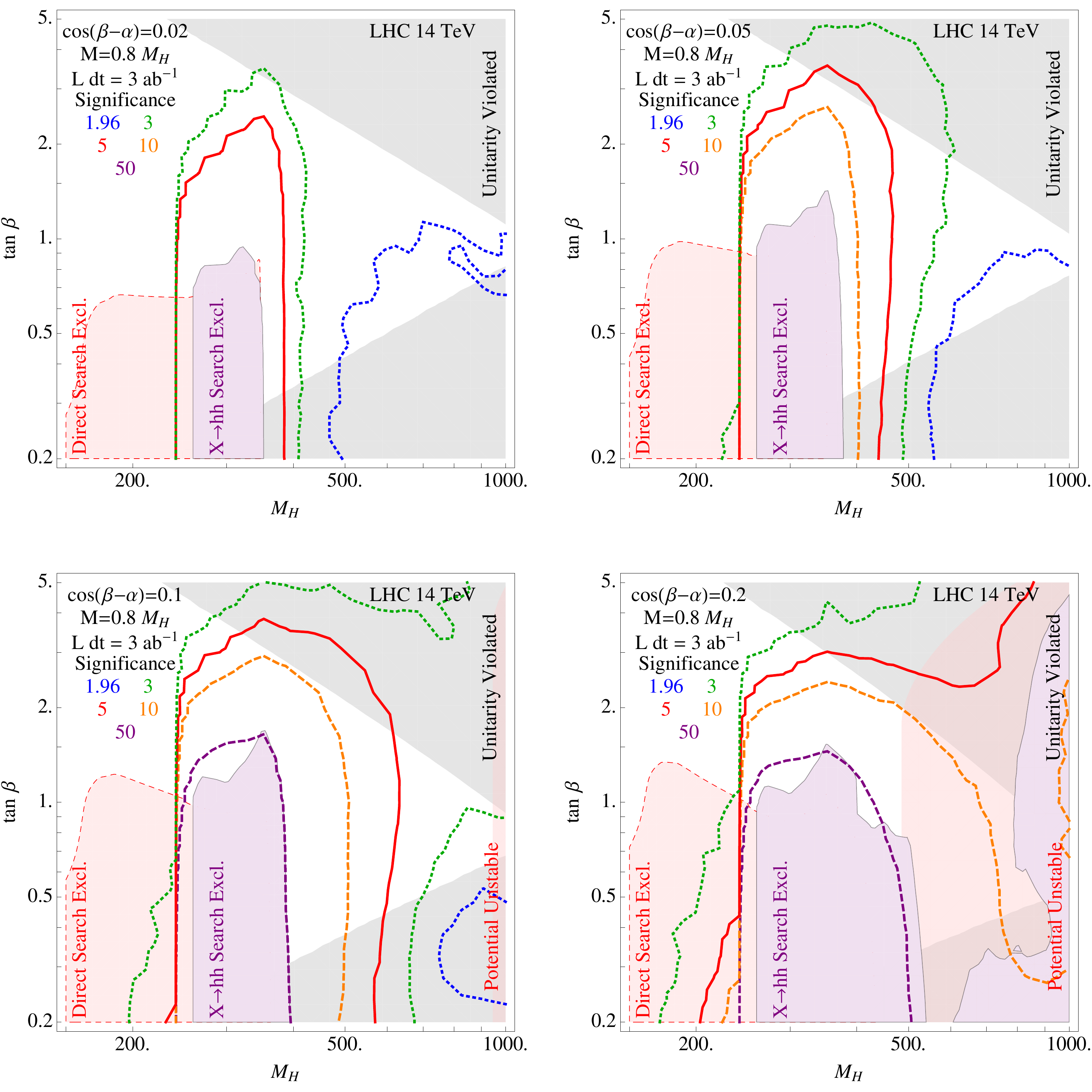}
\caption{Contours of the statistical significance with 3 ab$^{-1}$ of integrated luminosity in the plane of $\tan \beta$ and $M_H$ for selected values of $\cos(\beta-\alpha)$.  Additional experimental and theoretical constraints are shown as in Fig.~\ref{fig:lhhH}. }
\label{fig:discovery}
\end{center}
\end{figure}
The statistical significance expected with 3 ab$^{-1}$ of integrated luminosity at the LHC is shown in Fig.~\ref{fig:discovery}.  A bulk of the parameter space above $M_H > 2 m_h$ can be excluded at the 95\% C.L., even near the decoupling limit.

\section{Associated $hH$ production}
\label{sect:hH}

Associated production of a light-heavy Higgs boson pair is a valuable complement to $H \to hh$ resonant production for measuring components of the scalar potential, see Table~\ref{table:couplings}.   This process, by virtue of the scalar coupling, $\lhHH$, is not suppressed in the decoupling limit as seen in Eq.~\ref{eq:lhHH2}.

\begin{table}[hbpt]
\begin{center}
\begin{tabular}{|c|ccc|}
\hline
Process & $\lhhh$ & $\lhhH$ & $\lhHH$\\
\hline
$pp\to hh$ (continuum) & \Checkmark & $\times$ & $\times$\\
$pp\to H\to hh$ & \Checkmark & \Checkmark & $\times$\\
$pp\to h^*/H^* \to hH$  & \Checkmark & \Checkmark & \Checkmark\\
\hline
Decoupling dependence & $\lhhh_{\rm SM}(1+{\cal O}(\cbashort^2))$ & ${\cal O}(\cbashort)$ & $(2M_H^2- 2 M^2+M_h^2)/v + {\cal O}(\cbashort)$\\
\hline
\end{tabular}
\end{center}
\caption{The Higgs pair production processes that are sensitive to the couplings among the CP-even states. For each scalar coupling, the leading term in the expansion in the decoupling parameter, $\cbashort=\cba$, is also shown.}
\label{table:couplings}
\end{table}%

Both the box and triangle diagrams shown in Fig.~\ref{fig:feynman-diags} can contribute to $hH$ production. For the triangle diagram, we can have either $h$ or $H$ in the $s$-channel. Unless $M_H > 2 m_t$ and $\tb$ is small, the width of the heavy Higgs is narrow, so there is usually no enhancement for the $H$ diagram by being slightly off-shell. Therefore, all three diagrams are relevant.

\begin{figure}[htbp]
\begin{center}
\includegraphics[width=0.89\textwidth]{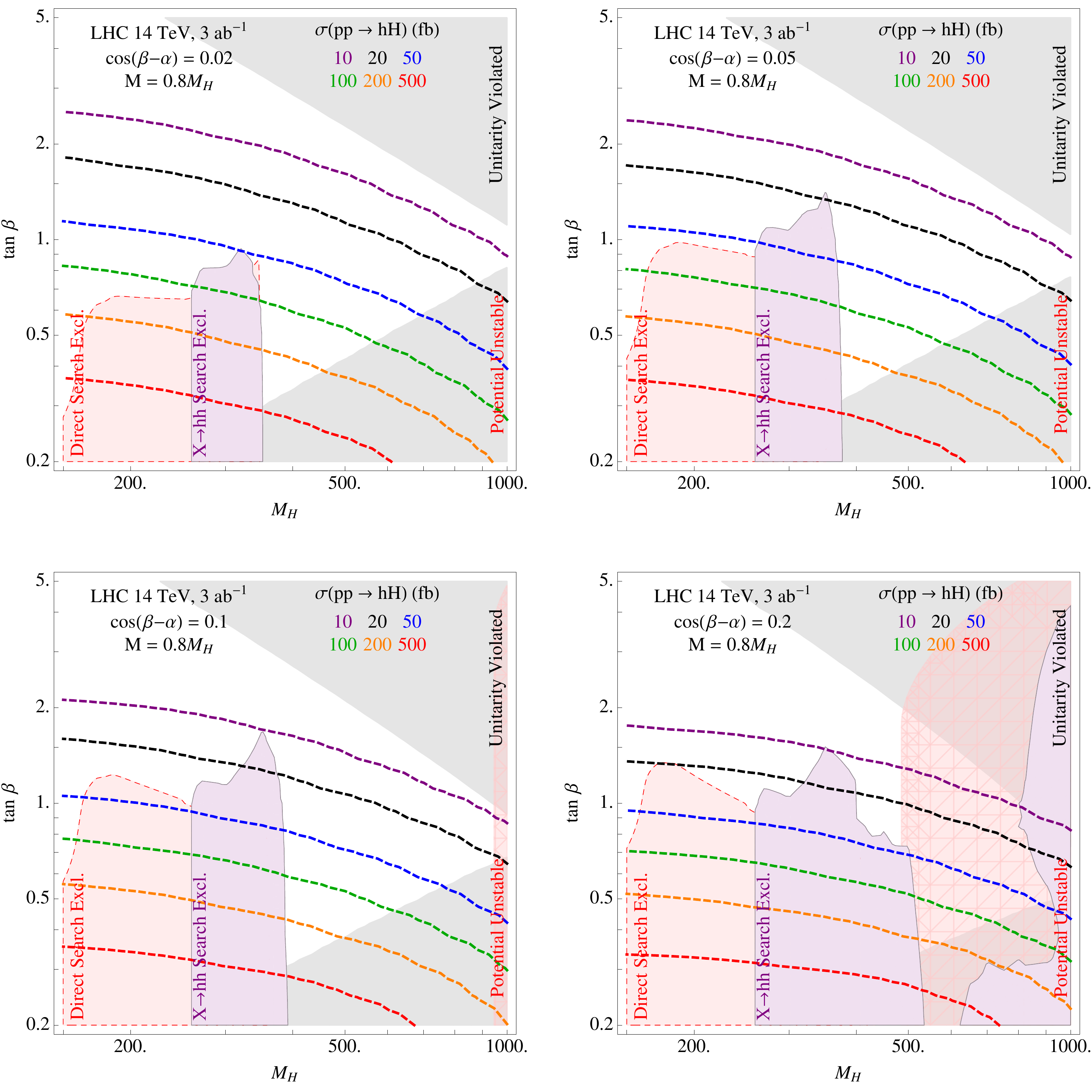}
\caption{Contours of $\sigma(pp\to hH)$ in the plane of $\tan \beta$ and $M_H$ for selected values of $\cos(\beta-\alpha)$. Additional experimental and theoretical constraints are shown as in Fig.~\ref{fig:lhhH}. }
\label{fig:hHxsec}
\end{center}
\end{figure}

As noted in Table~\ref{table:couplings}, the triangle diagram involving $H$ is the only Higgs pair process that probes $\lhHH$, as its amplitude is proportional to the combination $y_t^H \lhHH$. Furthermore, the contribution from the $h^*$ triangle diagram depends on the coupling combination $y_t^h \lhhH$, providing sensitivity to $\lhhH$, even in regions where $BF(H\to hh)$ is small.   The sensitivity is best for small $\tb$, due to the effect of the top Yukawa coupling to $H$ on the production cross section. More precisely, the magnitude of $y_t^H$ is largest at small $\tb$, as
\be
\frac{y_t^H}{y_t^{h_{SM}}} = c_{\beta-\alpha} - {\sqrt{1-c_{\beta-\alpha}^2}\over \tb.}
\ee
The production cross section for the $pp \to hH$ process is shown in Fig.~\ref{fig:hHxsec}.

The $hH$ process can proceed to a number of final states. As above, we let the light Higgs decay to either $\gamma \gamma$ or $\bb$. The preferred final state for $H$ depends strongly on $M_H$, and to a lesser extent $\cba$ and $\tb$. For $M_H<2 m_h$, $H$ decays predominantly into $b\bar{b}$ or $WW^{(*)}/ZZ^{(*)}$. $b\bar{b}$ is strongest for small $\cba$ and large $\tb$, while $WW^{(*)}/ZZ^{(*)}$ is most important for large $\cba$, as is demonstrated by the branching fraction contours shown in Fig.~\ref{fig:H2bb} and \ref{fig:H2VV} in Appendix~\ref{apx:bfs}, respectively.  For the $H \to \bb$ channel, we find that both the $4b$ and $b\bar{b}\,\gamma\gamma$ channels are viable.  We also explore $H \to ZZ \to 4\ell$ decays. We choose $ZZ\to 4l$ despite its small branching ratio because it has small backgrounds and allows for straightforward event reconstruction. However, this limits us to choosing $h\to \bb$ in order to have a detectable number of events at the LHC. 

Above $2 m_t$, $H$ decays primarily to $t$-quarks, with a branching fraction that surpasses 90\% for small $\tb$ (see Fig.~\ref{fig:H2tt} in Appendix~\ref{apx:bfs}). The most viable channel in this region is $t\bar{t}\,b\bar{b}$, with at least one of the tops decaying leptonically to reject background. Between $2 m_h$ and $2 m_t$, the $hH \to hhh \to 4b\,\gamma \gamma$ is important, as we would expect from the results of the resonant $H$ production analysis.  The $H \to \bb$ and $H \to ZZ$ channels are weaker but are possibly still viable in this region as well.



\begin{figure}[hbtp]
\begin{center}
\includegraphics[width=0.99\textwidth]{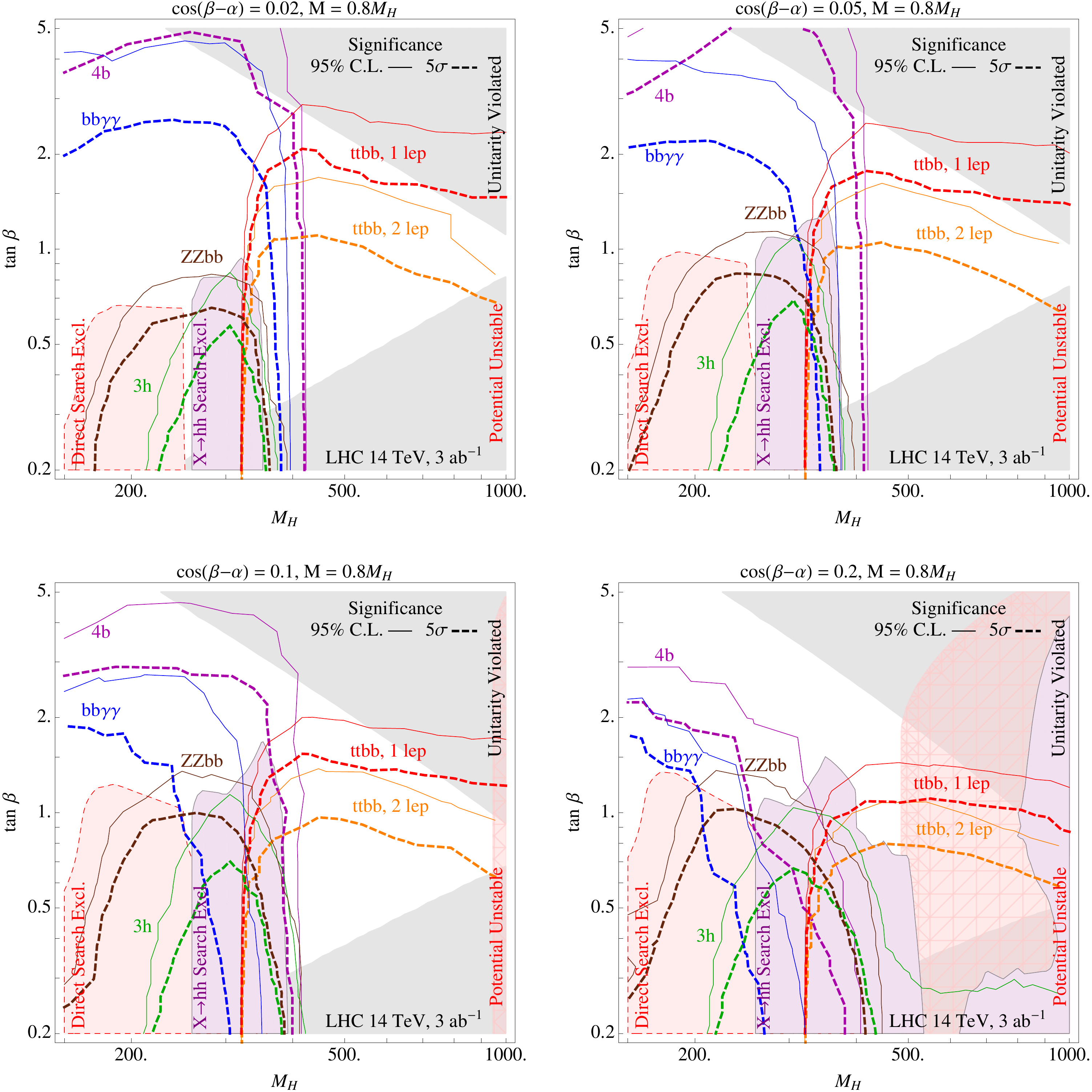}
\caption{Contours of the statistical significance with 3 ab$^{-1}$ of integrated luminosity in the plane of $\tan \beta$ and $M_H$ for selected values of $\cos(\beta-\alpha)$.   The bold dashed curves show the expected $5 \sigma$ significance, while the thin solid curves show the expected 95\% C.L. reach. The colors correspond to different final states: $\bbaa$ (blue), $4b$ (purple), $ZZ\bb$ (brown), $3h$ (green), $\tt\bb$ in the single-lepton channel (red), and $\tt\bb$ with two leptons (orange).  Additional experimental and theoretical constraints are shown as in Fig.~\ref{fig:lhhH}.   }
\label{fig:hHsig}
\end{center}
\end{figure}

We simulate the $pp \to hH$ signal using MADGRAPH as described in Section \ref{sec:hhsim} and compute the expected LHC reach for 3 ab$^{-1}$ at 14 TeV. In Fig.~\ref{fig:hHsig}, we show the expected $95\%$ CL and $5\sigma$ contours for the $\bbaa$, $4b$, $ZZ\bb$ $4b\gamma\gamma$, $t\bar{t}b\bar{b}$ (1 lepton), and $t\bar{t}b\bar{b}$ (2 lepton) final states. As the coupling $\lhHH$ is not suppressed in the decoupling limit, we find that our sensitivity is actually best for small $\cba$.  Indeed, for the smallest values of $\cba$, we find that LHC will be able to probe essentially all of the allowed parameter space at the $95\%$ CL.  Even for larger values of $\cba$, the LHC will be sensitive to up to $\tb\sim 2$ over a wide range of $M_H$.

In the following sections, we describe our background simulations and selection cuts for each channel. 
Throughout the analysis, we use the efficiencies and fake rates described in Section \ref{sect:bbaa}.
We also include a lepton to photon fake rate of $\epsilon_{e\to \gamma} =6.2\%$  \cite{Aad:2012tba}. 
Additionally, we apply the baseline cuts on $\Delta R$ and $p_T$ from the resonant $\bbaa$ analysis to all five channels, and add the following cuts for leptons: $\Delta R_{ab}> 0.2$, $p_T(\ell)> 20$ GeV, and $|\eta_{\ell}| < 2.4$.

\subsection{The $hH\to b \bar{b} \gamma \gamma$ channel}

The low-mass region is probed by the $H\to b\bar{b}$ channel. First let us consider the case where $h\to \gamma \gamma$. The $h\to \gamma \gamma$ branching fraction is extremely small ($2.3\times10^{-3}$), but requiring photons in the final state also reduces the background significantly. The irreducible backgrounds include
\bea 
pp &\to& \bbaa \\
pp &\to& b\bar{b} h \to b\bar{b} \gamma \gamma \\
pp &\to& Zh \to b\bar{b} \gamma \gamma,
\eea
while the reducible backgrounds are
\bea
pp &\to& b\bar{b}~e^+ e^-  \quad (e \to \gamma)\\
pp &\to& b\bar{b} j \gamma \quad~(j \to \gamma)  \\
pp &\to& b\bar{b} j j \\
pp &\to& j j \gamma \gamma \\
pp &\to& 3j+\gamma \\
pp &\to& 4j \text{ (negligible)}.
\eea
We require exactly two photons and two jets, with both jets $b$-tagged. Then we apply a multivariate analysis after incorporating the following basic cuts:
\be
|M_{\gamma\gamma} - m_h| < 5~{\rm GeV}, \quad M_{b\bar b} > 100~{\rm GeV}.
\ee
The first of these cuts isolates the light Higgs resonance, while the second rejects $Z/\gamma^* \to b\bar{b}$, as well as a significant portion of the continuum background.

From here, the procedure is exactly as it was in the resonant case discussed in Section \ref{sec:hhsim}.  Specifically, we form over a discriminant
\be
{\cal O}=\left\{ M_{\bbaa}, M_{b\bar b}, M_{\gamma\gamma}, p_T(b\bar b), p_T(\gamma\gamma), \Delta R_{b\bar b},\Delta R_{\gamma\gamma}\right\}.
\ee
The expected LHC significance for $\mathcal{L} = 3~\iab$ is shown in blue in Fig.~\ref{fig:hHsig}. The reach is extremely good for $M_H < 400$ GeV, especially for small $\cba$, where we achieve 95\% C.L. significance beyond $\tb \approx 5$. The cross section falls of rather quickly with $\cba$ due to the $\cba$ and $\tb$ dependence of the bottom Yukawa coupling to $H$, which is
\be
\frac{y_b^H}{y_b^{h_{\rm SM}}} =  c_{\beta - \alpha} + \sqrt{1-c_{\beta - \alpha}^2} \tb
\ee
Hence, $y_b^H$ is enhanced for large $\tb$, and this effect is strongest for small $\cba$.

\subsection{The $hH\to 4b$ channel}
We also consider the case where $H\to b\bar{b}$, but the light Higgs decays to $b\bar{b}$ instead of $\gamma\gamma$. The signal is much larger than in the $\bbaa$ case, but the QCD background is large as well. With appropriate cuts, we find that the two channels are comparable in significance. The irreducible backgrounds in this case are given by
\bea 
pp &\to& 4b \\
pp &\to& b\bar{b} h \to 4 b.
\eea
The main reducible background is $pp \to b\bar{b} j j$, with the jets faking $b$ quarks. We also considered the $4j$, $ t\bar{t}b\bar{b}$, and $Zh$, and $Wh$ backgrounds, but found them to be negligible.

\begin{figure}[hbtp]
\begin{center}
\includegraphics[width=0.63\textwidth]{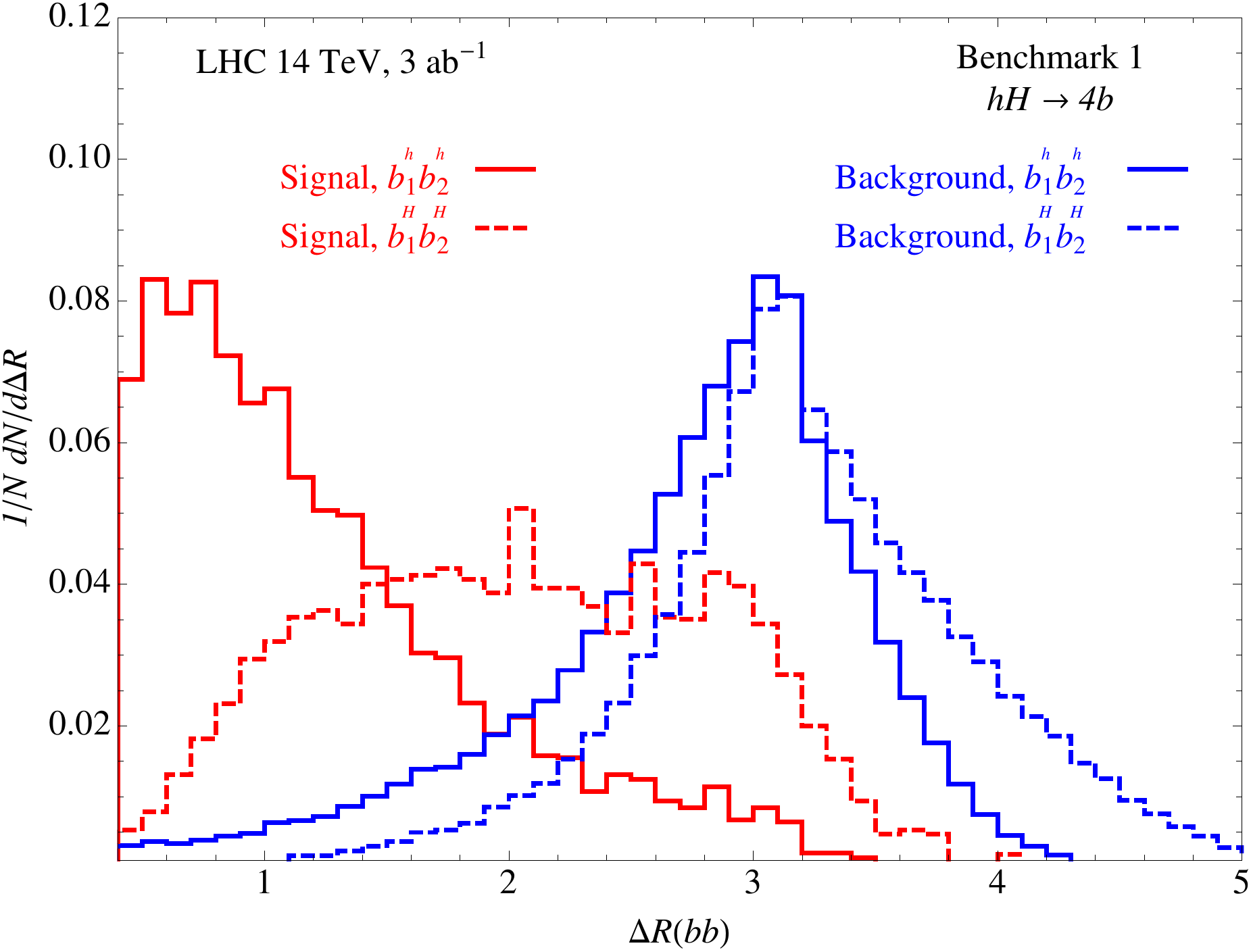}
\caption{The normalized $\Delta R$ distributions for the reconstructed light Higgs $bb$ pair (solid curves) and heavy Higgs $bb$ pair (dashed curves) are shown. The signal distributions for Benchmark 1 ($M_H = 300$ GeV, $\tb = 2$, and $\cba = 0.1$) are shown in red, while the background is shown in blue. The signal events tend towards smaller values of $\Delta R$, indicating that the Higgs bosons (especially the $h$) are boosted.}
\label{fig:deltaR}
\end{center}
\end{figure}

\begin{figure}[hbtp]
\begin{center}
\includegraphics[width=0.63\textwidth]{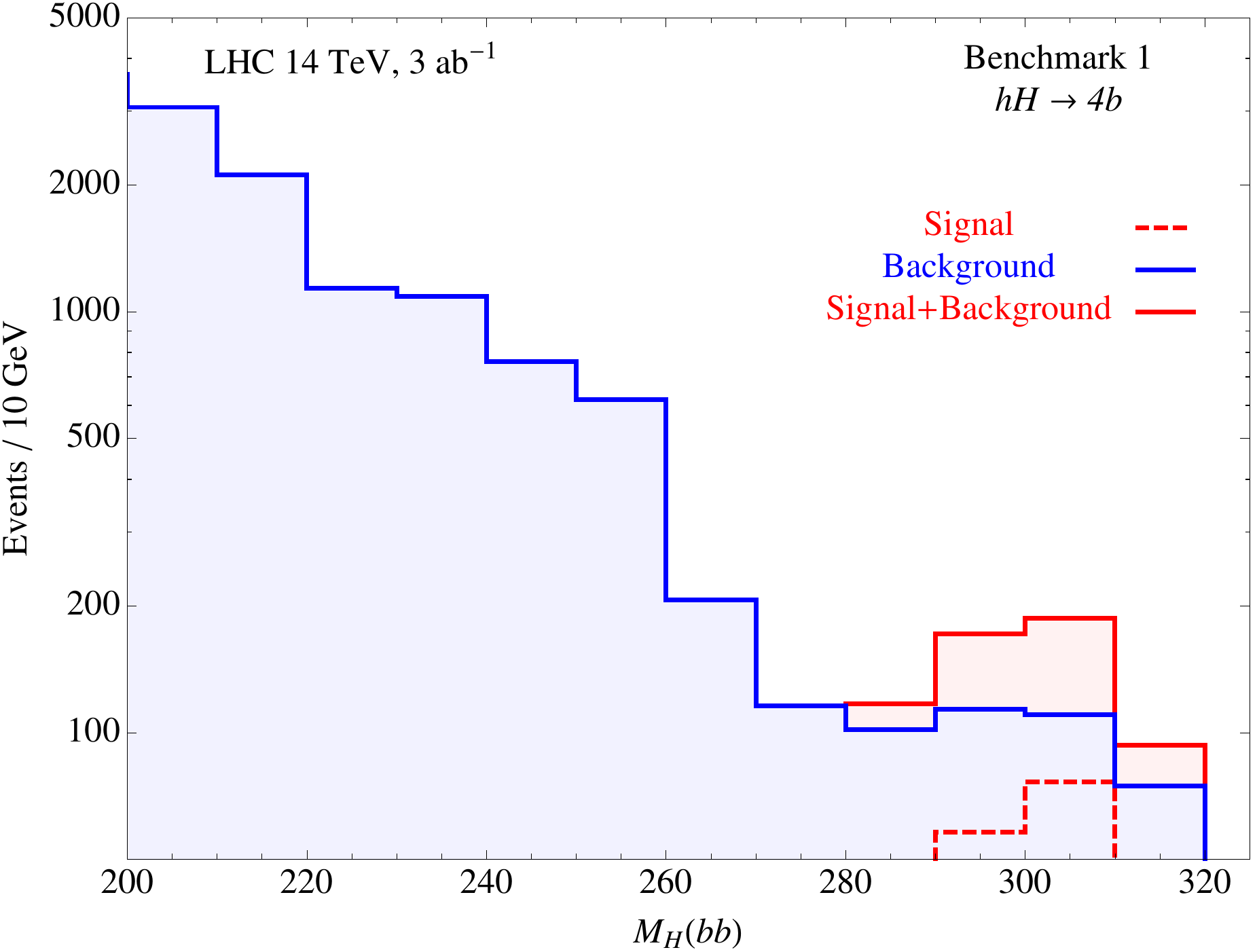}
\caption{The invariant mass of the heavier $bb$ pair in the $4b$ channel after the cuts on $M_h$ and $\Delta R$ is shown. Signal (red) is shown for Benchmark 1 ($M_H = 300$ GeV, $\tb = 2$, and $\cba = 0.1$). The background distribution (blue) drops off quickly with $M(bb)$, which leaves the resonance at $M_H$ clearly discernible.}
\label{fig:mHbb}
\end{center}
\end{figure}

We use a cut-based analysis. We require exactly four jets, all $b$-tagged. While the $b$-tagging efficiency is low, we find that all four $b$-tags are necessary to sufficiently reduce the light jet backgrounds. Since we have more than two $b$ quarks in the final state, care must be taken in reconstructing the parent Higgs bosons. We identify the decay products of the light Higgs by minimizing $|M_{b_i,b_j}-m_h|$ over all possible pairs $b_i,b_j$; we label the resulting pair as $b^h_1$ and $b^h_2$. The remaining two $b$ quarks are taken to reconstruct the heavy Higgs, and are labeled $b^H_1$ and $b^H_2$. After identifying the $b$ quarks, we apply the following cuts:
\bea
M(b^H_1 b^H_2) &>& 100~{\rm GeV}  \\
|M(b^h_1 b^h_2) - m_h| &<& 12.5~{\rm GeV}  \\
\Delta R(b^h_1,b^h_2)&<&1 \\
\Delta R(b^H_1,b^H_2)&<&1.5\\
|M(b^H_1 b^H_2) - M_H| &<& 15~{\rm GeV} . 
\eea
The $\Delta R$ cuts help isolate the signal from background. The light Higgs recoils against the heavy Higgs, so the two tend to have large $p_T$ and be well-separated in the $\phi-\eta$ plane. Since $M_H > m_h \gg m_b$, the Higgs decay products tend to be cluster, especially for the light Higgs. Therefore the $\Delta R$ distributions will be peaked at small values. Furthermore, the $\Delta R$ cuts also improve the reconstruction of $M_H$ by ensuring the $b$ quarks have been correctly paired. In Fig.~\ref{fig:deltaR}, we show the normalized $\Delta R$ distributions for both pair of $b$ quarks for $M_H = 300$ GeV, $\tb = 2$, and $\cba = 0.1$ before cuts. In Fig.~\ref{fig:mHbb}, we show the invariant mass distribution for the reconstructed heavy Higgs after the $\Delta R$ cuts for the same benchmark point.

The expected LHC significance for $\mathcal{L} = 3~\iab$ is shown in purple in Fig.~\ref{fig:hHsig}. The $4b$ channel is slightly stronger than the $\bbaa$ channel. As in the $\bbaa$ channel, the reach at large $\tb$ is good due to the high $BF(H\to \bb)$ in that region.

\subsection{The $hH\to ZZ \bb$ channel}

A complementary channel in the low-mass region is $H\to ZZ \to 4 \ell$,  $h\to b\bar{b}$. The only significant background is 
\be
pp \to t\bar{t} Z, 
\ee
with the tops decaying leptonically. The potential $ZZh$, $ZZjj$ and $WWZ$ backgrounds are negligible. There are also contributions to the signal from $h\to Z Z^*$, $H\to b\bar{b}$, but they are subdominant except for a small region with $M_H \lesssim 200$GeV, $\cba \lesssim 0.02$ and $\tb\gtrsim 5$. Furthermore, the resonant peaks in $M_{b\bar{b}}$ and $M_{4\ell}$ are well-separated between the two cases, so there is little interference. 

We require four leptons and two $b$-tags in our final state, then use the MVA to isolate the signal from the background. Our MVA variables are
\be
{\cal O}=\left\{ M_{4\ell}, M_{b\bar b},  p_T(b\bar b), \Delta R_{b\bar b}, \,\slash\!\!\!\!E_T\right\}.
\ee
The missing transverse energy variable is particularly important in this case, since $t\bar{t}Z$ has the same visible particle content as $ZZ b\bar{b}$, but with missing energy from the $W\to \ell \nu$ decays. This channel is promising for moderate values of $M_H$, especially for larger $\cba$. The LHC 3 ab$^{-1}$ significance is shown in brown in Fig.~\ref{fig:hHsig}.

\subsection{The $hH\to hhh \to 4b \gamma \gamma$ channel}
The $4b \gamma \gamma$ backgrounds are fairly small. The relevant backgrounds are
\bea 
pp &\to& 4b \gamma \gamma\\
pp &\to& b\bar{b} j j \gamma \gamma.
\eea
Backgrounds with higher light jet multiplicities are negligible due to the small $j\to\gamma$ and $j \to b$ fake rates. 

Our initial selection requires exactly four $b$-tagged jets and two photons. The diphoton invariant mass must satisfy
\be
|M_{\gamma\gamma} - m_h| < 5~{\rm GeV}.
\ee
This cut is sufficient to optimize the cut-based significance, reducing the background to only $\sim2$ events for 3~ab$^{-1}$. However, we can better reconstruct the heavy Higgs mass with an additional cut. First we pair the $b$ quarks by minimizing $|M_{b_i,b_j}-M_h|$ over all possible pairs $b_i,b_j$, as in the $4b$ final state. We denote the three reconstructed light Higgs bosons as ($h_\gamma, h_{b1}, h_{b2})$, where $|M(h_{b1})-M_h|<|M(h_{b2})-M_h|$.  
If we compute $\Delta R(h_i,h_j)$ for each of the reconstructed light Higgs bosons, we find that the distribution is peaked at low $\Delta R$ and near $\Delta R \approx \pi$.  This corresponds to two light Higgs bosons from the $H$ decay being clustered together and the other $h$ recoiling against the $H \to hh$ system.  We therefore require that exactly one pair $(h_i,h_j)$ satisfy
\be
\Delta R(h_i,h_j) < 1.5,
\ee
and use that pair to reconstruct the heavy Higgs. 

The expected LHC significance for $\mathcal{L} = 3~\iab$ is shown in green in Fig.~\ref{fig:hHsig}. Unsurprisingly, the significance contours run parallel those found for the $H\to hh$ resonant case.

\subsection{The $hH\to b\bar b t \bar t$ channel}
To explore the $M_H \gtrsim 2 m_t$ region, we consider the $b\bar b t \bar t$ final state. The irreducible backgrounds include:
\bea 
pp &\to& b\bar{b}t\bar{t} \\
pp &\to&h  t\bar{t}  \to b\bar{b} t\bar{t} \\
pp &\to&  Z  t\bar{t}  \to  b\bar{b} t\bar{t}.
\eea
We also include the reducible background $pp \to jjt\bar{t}$. 

We require four $b$-tags in our final state. At least one of the top quarks must decay via $t \to b W \to b\ell \nu$. We allow the other top to decay to $b j j$ or $b\ell \nu$. Thus our final state must include either $4j+2\ell+\slash \!\!\!\!E_T$ or $6j+\ell+\slash \!\!\!\!E_T$. As above, we reconstruct the light Higgs by minimizing $|M_{b_i,b_j}-m_h|$ and requiring that this pair of $b$ quarks satisfies
\be
|M(b^h_1 b^h_2) - m_h| < 12.5~{\rm GeV}\\
\ee
and
\be
\Delta R(b^h_1 b^h_2)  < 1.0.
\ee
The other two $b$ quarks are assumed to come from top decays. The $\Delta R(b^h_1 b^h_2) $ is very peaked in this channel, since $H$ must be heavy to allow for $\tt$ decays. This leads to a more boosted light Higgs than in the previous channels, and therefore more closely clustered $b$ quarks. 

\begin{figure}[t!]
\begin{center}
\includegraphics[width=0.63\textwidth]{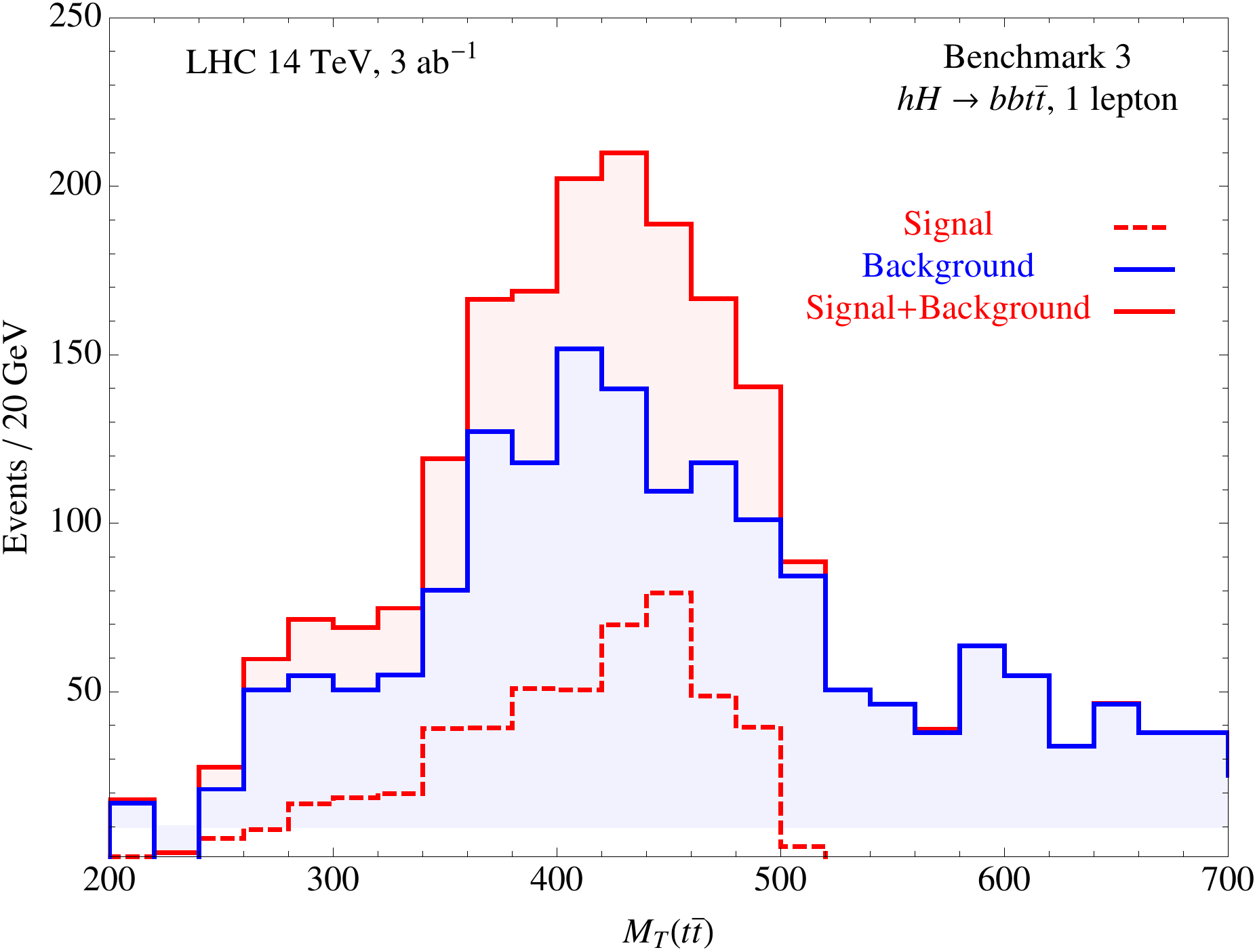}
\caption{The transverse mass of the top quark pair in the $\tt\bb$ with one lepton channel. The signal (red) is shown for Benchmark 3 ($\cba = 0.02$, $\tb = 1$, and  $M_H = 500$ GeV). The signal drops off sharply above $M_H$, while the background (blue) decreases more gradually. }
\label{fig:mTH}
\end{center}
\end{figure}
In the one-lepton channel, we apply additional cuts. We can reconstruct the tops by minimizing  $|M(b_i jj)-m_t|$ over the remaining two $b$ quarks. Let $M(t_h) = M(bjj)$ and $M_T(t_l) = M_T(b\,l\, \slash \!\!\!\!E_T) $. Then we require
\be
|M(t_h) -m_t | < 20 ~{\rm GeV}
\ee
and
\be
M_T(t_l) < m_t.
\ee
Finally, we define a signal region that varies with $M_H$:
\be
M_H - 200 {\rm GeV} < M_T(t_h t_l) < M_H - 10 {\rm GeV.}
\ee
In Fig.~\ref{fig:mTH}, we show the transverse mass of the top quark pair after cuts for $\cba = 0.02$, $\tb = 1$, and  $M_H = 500$ GeV.  The one-lepton channel is stronger than the two-lepton channel due to the relatively small branching fraction for $W\to \ell \nu$.  The expected LHC significances for $\mathcal{L} = 3~\iab$ are shown in red and orange in Fig.~\ref{fig:hHsig}. The reach decreases slowly with $M_H$, and it should be possible to probe above $M_H = 1~{\rm TeV}$ for $\tb<2$ at the LHC.

\section{Conclusions}
\label{sec:conclusion}

We have investigated two types of Higgs pair production within the CP-conserving Type-II 2HDM: the resonant production of an $hh$ pair, and the associated production of an $hH$ pair.  We included theoretical constraints from requiring perturbative unitarity and a bounded scalar potential, as well as LHC constraints from the direct heavy Higgs search and the $X\to hh$ search. We have made the simplifying assumptions that $M_H = M_{H\pm} = M_A $ ,and $M= 0.8 M_H$ and have presented our results in terms of the remaining free parameters: $M_H$, $\tb$, and $\cba$.

For the resonant case of $pp\to H\to hh$, the reach in the $\bbaa$ channel for 30 fb$^{-1}$ at LHC14 is comparable to the current limits on $X\to hh$, as expected. 
With 3 ab$^{-1}$, the coverage extends to $\tb \approx 2$ and $M_H \approx 350$ GeV near the decoupling limit.  For large $\tb$, the reach improves so that a majority of the theoretically allowed region above $M_H = 2 m_h$ may be probed. This is because the $H\to hh$ rate is governed by the $\lhhH$ coupling, which behaves as $\cba$ to leading order and is suppressed in the decoupling limit. 

The associated production case, $pp\to hH$, offers a variety of interesting channels to explore.  Near the decoupling limit, the LHC14 reach is excellent due to the non-decoupling nature of the $\lhhH$ scalar coupling. Due to the potentially large mass difference between light and heavy Higgs states,  the $h$ is often boosted when $M_H \gg m_h$, resulting in decay products which have small separation.  This is contrary to the common backgrounds, which contain more dispersed jets and leptons, resulting in a quite clean differentiation between signal from background.  In the low mass region, $M_H < 2 m_h \simeq 250$ GeV, the $H\to bb, h\to \bb/\gamma\gamma$ channels cover the entire allowed range of $\tb$.  The $Hh\to \tt\bb$ channels cover the high mass region, $M_H > 2 m_t \simeq 350$ GeV. For larger values of $\cba$, the sensitivity in these channels decreases due to the increased $BF(H\to WW/ZZ)$, when kinematically allowed.  However, $H\to ZZ$ and $H\to hh$ improve the reach in this region.

In our analysis, we selected three benchmark points that illustrate the discovery potential for different channels, which were presented in Table~\ref{tab:bench}.  Point A, for which $M_H = 300$ GeV, $\tb = 2$, and $\cba = 0.1$, demonstrated the viability of the $H\to hh\to \bbaa$ channel due to the large BF$(H\to hh)$.  A secondary channel that is viable is the $Hh\to b\bar b b\bar b$ mode.  Point B, for which $M_H = 300$ GeV, $\tb = 1$, and $\cba = 0.02$,  highlighted the $hh/hH\to \bbaa$ and $b\bar b b\bar b$ channels. The large BF$(H\to \bb)$ provides a sizable rate to the $\bbaa$ and $\bb\bb$ final states.  The $hh\to \bbaa$ channel has high significance due to the large production cross section of $pp\to hh$.  Point C, for which $M_H = 500$ GeV, $\tb = 1$, and $\cba = 0.02$, highlighted the $hH\to t\bar t b\bar b$ channel.  In this case,  BF$(H\to t\bar t)$ is large, allowing a sizable rate for the final state.  We present the statistical significance for these points at LHC14 with 3 ab$^{-1}$ of integrated luminosity in Table~\ref{table:benchmarks}.

\begin{table}[t]
\renewcommand{\arraystretch}{1.33}
\begin{center}
\begin{tabular}{|c |c |c| c|}
\cline{2-4}
\multicolumn{1}{ c| }{} & {\bf A}:&{\bf B}:&{\bf C}: \\[-1.5mm]
  \multicolumn{1}{ c| }{} & $M_H = 300$ GeV,&$M_H = 300$ GeV,& $M_H = 500$ GeV, \\[-2mm]
 \multicolumn{1}{ c| }{} &$t_\beta = 2$, $c_{\beta -\alpha}= 0.1$&$t_\beta = 1$, $c_{\beta -\alpha} = 0.02$& $t_\beta = 1$, $c_{\beta -\alpha}= 0.02$\\
 \hline
 $\lhhh / \lhhh_{SM}$  &0.946&0.998&0.992 \\
$\lhhH$ (GeV) &40.8&8.87&29.2 \\
$\lhHH$ (GeV) &310&327&795\\
$y^H_t$ &$-0.40$&$-0.98$&$-0.98$\\
   \hline
$\sigma(pp \to hh)$ (fb)& 340 &810 & 37 \\
$\sigma(pp\to hH)$ (fb)& 7.7 & 44 & 26 \\
\hline
$BF(H\to hh)$&18\%&7.6\%&0.1\% \\
$BF(H\to tt)$&0.0\%&0.0\%&99\% \\
$BF(H\to bb)$&34\%&74\%&0.2\% \\
$BF(H\to ZZ+WW)$&49\%&18\%&0.2\% \\
\hline
${\cal S}(H\to hh)$  &22 &55&2.4 \\
${\cal S}(Hh \to 3h \to  b\bar b b\bar b \gamma \gamma)$  &0.38&1.2&0.0 \\
${\cal S}(Hh\to \bbaa )$  &2.5&14&0.0 \\
${\cal S}(Hh\to  b\bar b b\bar b) $ &8.2&68& 0.0\\
${\cal S}(Hh\to t_h t_\ell \bb) $   &0.0&0.0&16\\
${\cal S}(Hh\to t_\ell t_\ell \bb) $  &0.0&0.0& 5.6\\
${\cal S}(Hh\to ZZ\bb) $ &0.62&0.48&0.0 \\
\hline
\end{tabular}
\end{center}
\caption{The three benchmark points chosen to help elucidate the most viable channels as in Table~\ref{tab:bench}, but with the expected statistical significance for LHC14 with 3 ab$^{-1}$ of integrated luminosity.}
\label{table:benchmarks}
\end{table}%

Ultimately, the results of our analysis demonstrate that there is a large region of the CP-conserving Type-II 2HDM parameter space that is currently unconstrained, but should be testable by the LHC 14 TeV run. Resonant production of $hh$ pairs and associated production of $hH$ pairs are orthogonal probes of the 2HDM scalar potential. By considering both production modes, along with the continuum production of $hh$ pairs, the LHC should be able to measure the three triscalar couplings ($\lhhh$, $\lhhH$, $\lhHH$). These coupling measurements can then be checked for consistency with a given model in order to illuminate the structure of the underlying scalar sector.


\section{Acknowledgements}
V.~B, L.~L.~E, A.~P. and G.~S. are  supported by the U. S. Department of Energy under the contract DE-FG-02-95ER40896. 

\appendix

\section{Heavy Higgs Branching Fractions}
\label{apx:bfs}
We calculate the branching fractions of the Heavy Higgs via the expected SM-like partial widths
\bea
\Gamma_{H\to VV} &=& \cba^2 \Gamma^{\rm SM}_{H\to VV},\\
\Gamma_{H\to \bb}&=& (y_t^H/y_t^{h_{\rm SM}})^2 \Gamma^{\rm SM}_{H\to \bb},\\
\Gamma_{H\to \tt}&=&  (y_b^H/y_b^{h_{\rm SM}})^2 \Gamma^{\rm SM}_{H\to \tt},\\
\Gamma_{H\to \tautau}&=&  (y_\tau^H/y_\tau^{h_{\rm SM}})^2 \Gamma^{\rm SM}_{H\to \tautau}.
\eea
where $\Gamma^{\rm SM}$ indicates the SM-like partial width with $M_{h_{\rm SM}} = M_H$.  For these calculations, we neglect the partial decays to $\gamma\gamma$ and $gg$ and light quarks as they're negligible for the cases we consider.  We calculate the SM-like Higgs partial widths with the HDECAY package~\cite{Djouadi:1997yw}.  The heavy Higgs partial width to the SM-like Higgs boson at $m_h=125$ GeV is given by
\be
\Gamma_{H\to hh} = {\left(\lhhH\right)^2 \over 32 \pi M_H} \sqrt{1-{4 m_h^2 \over M_H^2}}.
\ee
\begin{figure}[h!]
\begin{center}
\includegraphics[width=0.99\textwidth]{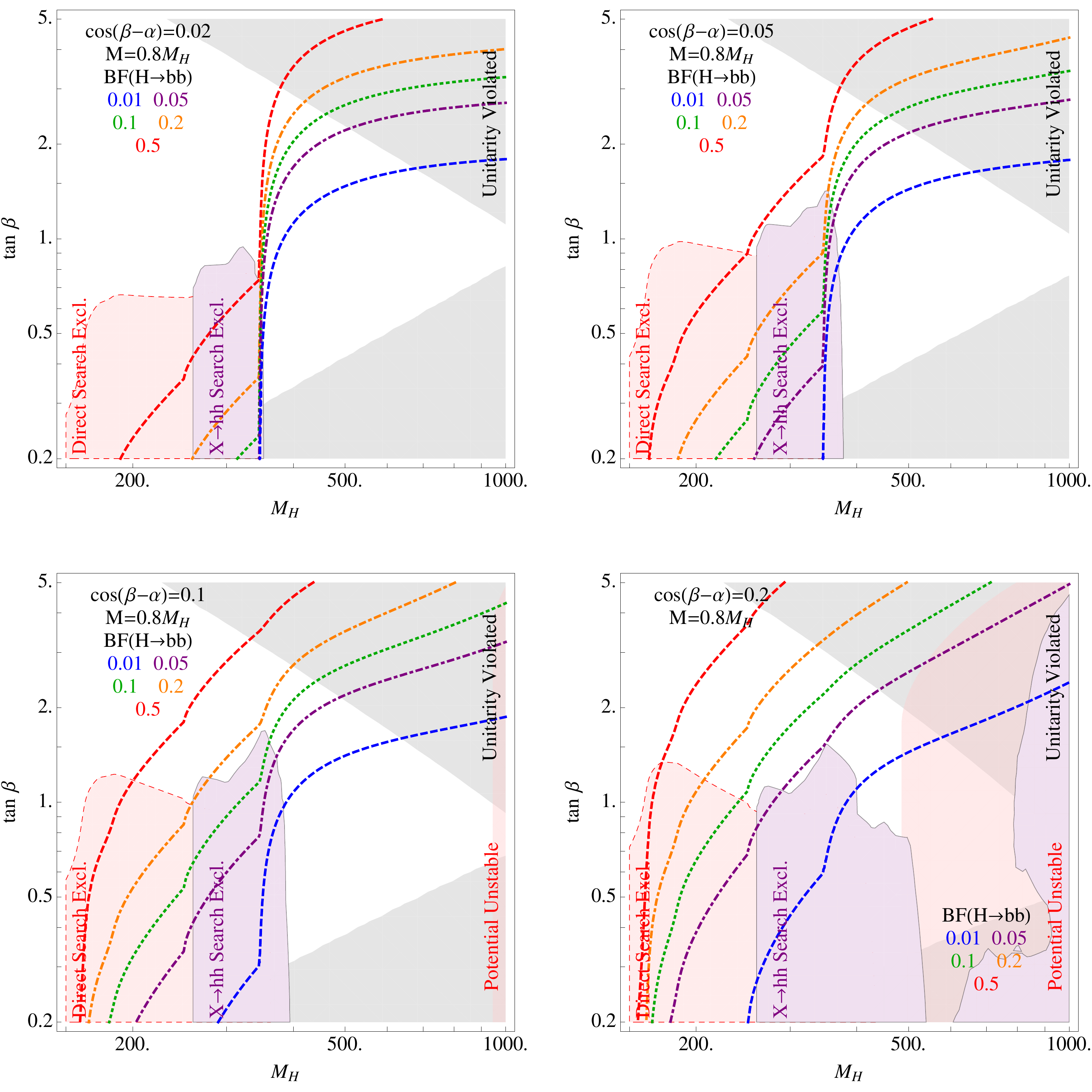}
\caption{Contours of $BF(H\to b\bar b)$ in the plane of $\tan \beta$ and $M_H$ for selected values of $\cos(\beta-\alpha)$. Additional experimental and theoretical constraints are shown as in Fig.~\ref{fig:lhhH}.   }
\label{fig:H2bb}
\end{center}
\end{figure}
\begin{figure}[h!]
\begin{center}
\includegraphics[width=0.99\textwidth]{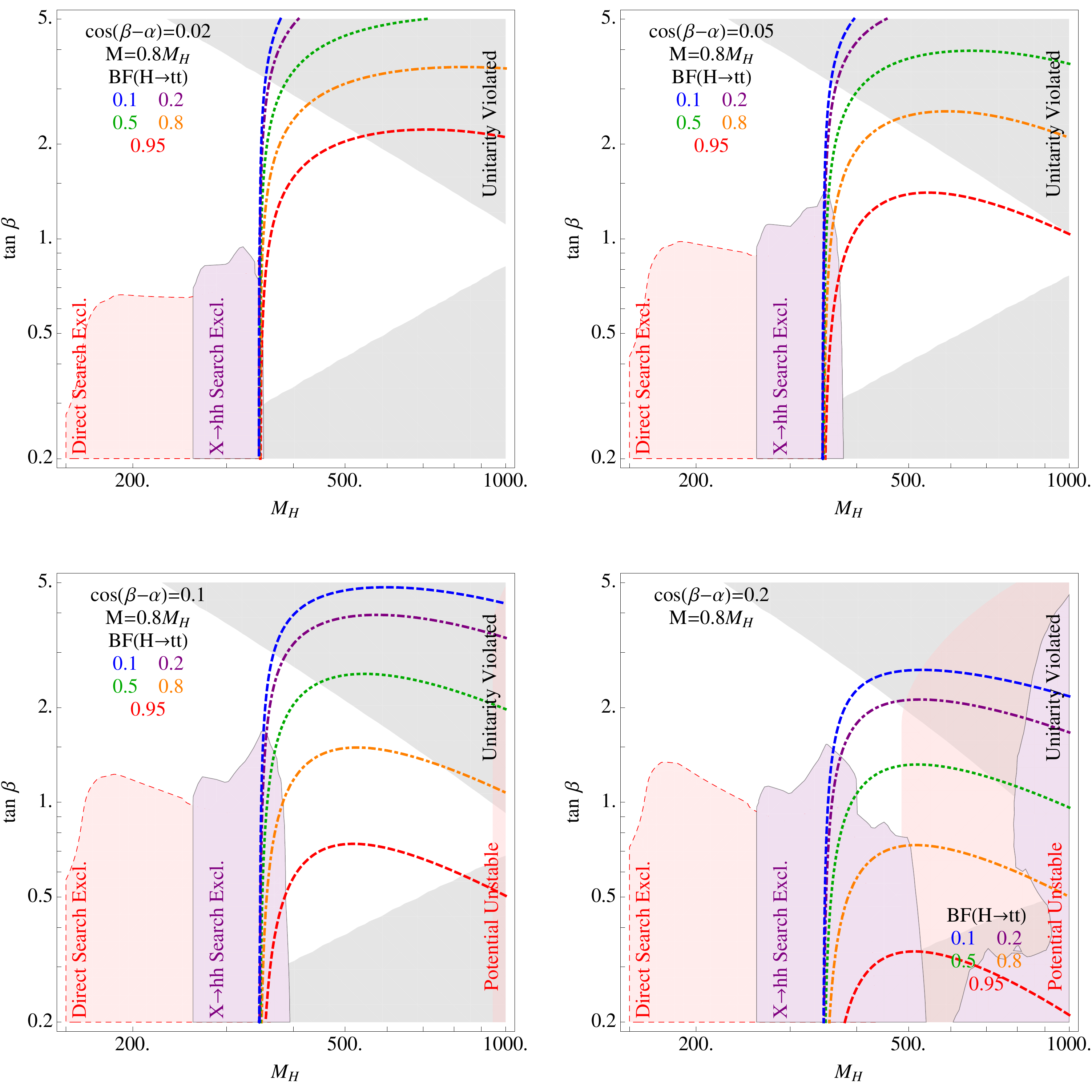}
\caption{Contours of $BF(H\to t\bar t)$ in the plane of $\tan \beta$ and $M_H$ for selected values of $\cos(\beta-\alpha)$.  Additional experimental and theoretical constraints are shown as in Fig.~\ref{fig:lhhH}.   }
\label{fig:H2tt}
\end{center}
\end{figure}
\begin{figure}[h!]
\begin{center}
\includegraphics[width=0.99\textwidth]{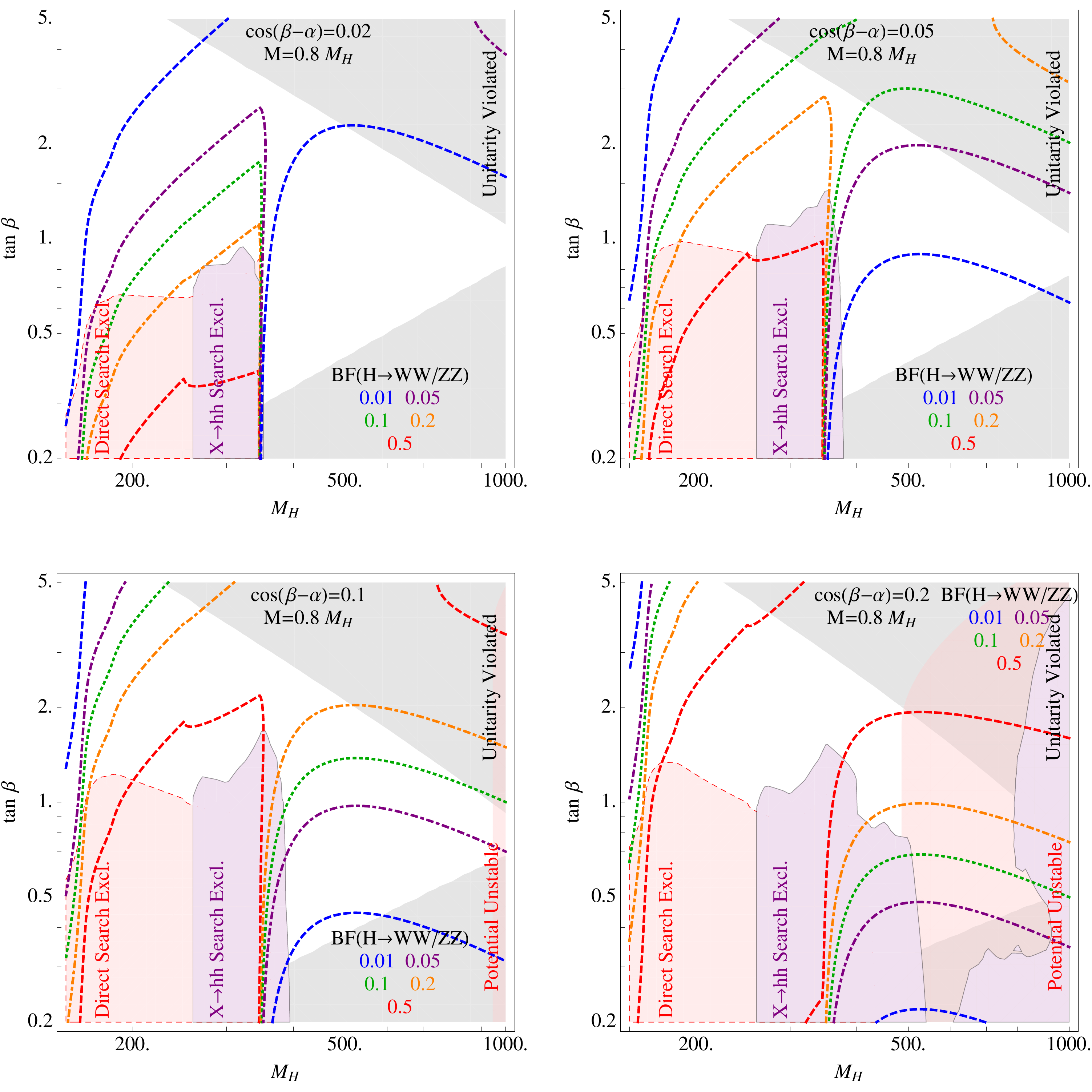}
\caption{Contours of $BF(H\to WW^{(*)}/ZZ^{(*)})$ in the plane of $\tan \beta$ and $M_H$ for selected values of $\cos(\beta-\alpha)$.   Additional experimental and theoretical constraints are shown as in Fig.~\ref{fig:lhhH}.   }
\label{fig:H2VV}
\end{center}
\end{figure}

The total Higgs boson width is calculated according to the sum of the respective partial widths
\be
\Gamma_H = \Gamma_{H\to VV} + \Gamma_{H\to \bb} + \Gamma_{H\to \tt} + \Gamma_{H\to \tautau} + \Gamma_{H\to h h}, 
\ee
leading to the branching fractions that are calculated in the usual way
\be
\text{BF}(H\to X\bar X) = {\Gamma_{H\to XX}\over\Gamma_H},
\ee
where $XX=VV,\bb,\tt,\tautau$ and $hh$.  We list in Figs.~\ref{fig:H2bb},~\ref{fig:H2tt}, and \ref{fig:H2VV} the contours of branching fractions in the selected parameter planes for $\bb, \tt$ and $VV$, respectively.  The branching fraction to $hh$ is shown in Fig.~\ref{fig:H2hh} in Section~\ref{sect:scalarcoup}.

 \end{document}